\documentclass[twocolumn,twocolappendix]{aastex7} 
\usepackage[utf8]{inputenc}
\usepackage{enumerate}
\usepackage{graphicx}
\usepackage{mathtools}
\usepackage{comment}
\usepackage{tabularx}
\usepackage{amsmath}
\usepackage{changepage}
\usepackage{amssymb}
\usepackage{lipsum} 
\usepackage{float}
\usepackage{commath}
\usepackage{enumerate}
\usepackage{longtable}
\usepackage{stackengine}
\usepackage{xcolor}
\usepackage{cleveref}
\usepackage[caption=false]{subfig}
\usepackage{placeins}
\usepackage[ruled,lined]{algorithm2e}

\AfterEndEnvironment{figure}{\noindent\ignorespaces}

\begin{document}
\title{Measurement of Substructure from the Kinematics of the GD-1 Stellar Stream} 
\author[0000-0001-8042-5794]{Jacob Nibauer}
\altaffiliation{NSF Graduate Research Fellow}
\affiliation{Department of Astrophysical Sciences, Princeton University, 4 Ivy Ln, Princeton, NJ 08544, USA}\email{jnibauer@princeton.edu}

\author[0000-0002-7846-9787]{Ana Bonaca}
\affiliation{The Observatories of the Carnegie Institution for Science, 813 Santa Barbara Street, Pasadena, CA 91101, USA}\email{abonaca@carnegiescience.edu}

\author[0000-0003-0872-7098]{Adrian M. Price-Whelan}
\affiliation{Center for Computational Astrophysics, Flatiron Institute, 162 5th Avenue, New York, NY, 10010, USA}\email{aprice-whelan@flatironinstitute.org}

\author[0000-0002-5151-0006]{David N. Spergel}
\affiliation{Center for Computational Astrophysics, Flatiron Institute, 162 5th Avenue, New York, NY, 10010, USA}
\affiliation{Department of Astrophysical Sciences, Princeton University, 4 Ivy Ln, Princeton, NJ 08544, USA}\email{president@simonsfoundation.org}

\author[0000-0002-5612-3427]{Jenny E. Greene}
\affiliation{Department of Astrophysical Sciences, Princeton University, 4 Ivy Ln, Princeton, NJ 08544, USA}\email{jgreene@astro.princeton.edu}

\correspondingauthor{Jacob Nibauer}
\email{jnibauer@princeton.edu}
\begin{abstract}
\noindent
Stellar streams are sensitive tracers of low-mass dark matter subhalos and provide a means to test the Cold Dark Matter (CDM) paradigm on small scales. In this work, we connect the intrinsic velocity dispersion of the GD-1 stream to the number density and internal structure of dark matter subhalos in the mass range $10^5$–$10^9~M_\odot$. We measure the radial velocity dispersion of GD-1 based on 160 identified member stars across four different spectroscopic catalogs. We use repeat observations of the same stars to constrain binarity. We find that the stream’s intrinsic radial velocity dispersion ranges from approximately $2$--$5~\textrm{km}~\textrm{s}^{-1}$ across its length. The region of GD-1 with the highest velocity dispersion represents a $4\sigma$ deviation from unperturbed stream models formed in a smooth Milky Way potential, which are substantially colder. We use perturbation theory to model the stream's velocity dispersion as a function of dark matter subhalo population parameters, including the number of low-mass subhalos in the Milky Way, the dark matter half-mode mass, and the mass-concentration relation of subhalos. We find that the observed velocity dispersion can be explained by numerous impacts with low-mass dark matter subhalos, or by a single impact with a very compact subhalo with $M \gtrsim 10^8~M_\odot$. Our constraint on the fraction of mass in subhalos is $f_{\mathrm{sub}} = 0.05^{+0.08}_{-0.03}$ (68\% confidence). In both scenarios, our model prefers subhalos that are more compact compared to CDM mass-size expectations. These results suggest a possible deviation from CDM at low subhalo masses.
\vspace{1cm}
\end{abstract}

\section{Introduction}
In the cold-dark-matter (CDM) paradigm, it is predicted that galaxies should contain a large amount of substructure in the form of dark matter subhalos (e.g., \citealt{1999ApJ...522...82K, 1999ApJ...524L..19M, 2008MNRAS.391.1685S}). While massive subhalos contain a stellar component, it is expected that below the galaxy formation threshold ($\approx 10^8~M_\odot$;  \citealt{galaxies7040081}) that subhalos should be completely dark, without any baryonic component. Under CDM, dark matter subhalos are expected from the scale of galaxies down to the mass scale of the Earth. If the dark matter particle is a thermal relic (i.e., warm dark matter;  \citealt{PhysRevLett.72.17}), there is a suppression of subhalos below $\approx 10^9~M_\odot$. If the dark matter is self-interacting, then the central density of core-collapsed subhalos can be significantly higher than CDM expectations (e.g., \citealt{2000PhRvL..84.3760S}). Therefore, measuring the abundance and central density of subhalos below the threshold of galaxy formation would provide a crucial test of the CDM paradigm and the nature of dark matter.

Because low-mass subhalos would lack a stellar component, methods to constrain their statistics are based on indirect gravitational tracers. Gravitational lensing is one avenue, where flux ratio anomalies can be used to constrain subhalos down to $\approx 10^7~M_\odot$ (e.g., \citealt{1998MNRAS.295..587M,Dalal_2002,2020MNRAS.492L..12G}). Stellar streams provide another  probe of dark matter substructure, as their morphologies and kinematics are sensitive to perturbations from low-mass subhalos (e.g., \citealt{2002ApJ...570..656J,2002MNRAS.332..915I,2009ApJ...705L.223C}). In this work we derive dark matter constraints from stellar streams using kinematic data.

Stellar streams result from the disruption of a globular cluster in the potential of a more massive host. Stars escape from the cluster primarily out of the Lagrange points, forming a smooth kinematically cold distribution of co-moving stars. Because of their well-ordered nature in the absence of perturbations, streams provide a test of substructure, particularly low-mass (i.e., $10^6~M_\odot$) dark matter subhalos. The effect of small-scale perturbations from subhalos is to impart gaps and impulsive velocity kicks to tidal tails (e.g., \citealt{2011ApJ...731...58Y,2012ApJ...748...20C, 2015MNRAS.450.1136E,2016MNRAS.457.3817S,2017MNRAS.466..628B,2024arXiv241213144A}). 
Over 100 streams have been detected around the Milky Way \citep[e.g.,][]{2025NewAR.10001713B}.

Perhaps the most well-studied stellar stream is GD-1 \citep{Grillmair_2006}. The stream spans over 80~\rm{deg} across the sky and has a morphology indicative of encounters with substructure. In particular, the stream contains a well-defined spur and gap component, which can be explained by an interaction with a compact $10^6~M_\odot$ dark matter subhalo \citep{2019ApJ...880...38B}. Recently it was shown that the high concentration of the subhalo required to match observations is consistent with gravothermal core-collapse due to dark matter self interactions \citep{gd1_core_collapse}. Indeed, stellar streams provide a promising probe of not only the number of dark matter subhalos, but also their internal properties since more compact subhalos typically induce larger perturbations. 

In this work we measure the intrinsic velocity dispersion of the GD-1 stream, and determine what population of dark matter subhalos is needed to match the observed velocity dispersion. We apply the perturbative modeling framework, \texttt{streamsculptor} \citep{2025ApJ...983...68N}, to generate numerous realizations of the stream as a function of the number and concentration of subhalos, and marginalize over a range of stream ages and progenitor masses. We cast our results in an empirical manner, so that they may be compared to any dark matter model.

The paper is organized as follows. In \S\ref{sec: data} we introduce the dataset and describe the velocity dispersion measurement. In \S\ref{sec: model} we provide the ingredients of our model. In \S\ref{sec: results} we present our results. In \S\ref{sec: discussion} we discuss our results, and in \S\ref{sec: summary_and_discuss} we conclude.

\section{Data and Velocity Dispersion Measurement}\label{sec: data}
Stars are selected from the data-driven GD-1 catalog from \citet{2025ApJ...980..253S} with membership probability $>75\%$. Radial velocities are obtained from four datasets: the Dark Energy Spectroscopic Instrument early data release (DESI; \citealt{2024AJ....168...58D,2024MNRAS.533.1012K, 2025ApJ...980...71V}), SDSS DR9 \citep{2012ApJS..203...21A, 2019ApJ...877...13H}, LAMOST DR8 \citep{2012RAA....12.1197C}\footnote{\url{http://www.lamost.org/dr8/}\label{lamost}}, and MMT Hectochelle \citep{2020ApJ...892L..37B}. We select stars falling within $30~\rm{km/s}$ of the mean radial velocity track of our unperturbed models \citep{2025ApJ...983...68N}. The result is 195 radial velocity measurements, of which 34 are repeat observations of the same star with two different instruments, and one star is observed by three instruments. The total number of unique stars is 160.

\begin{figure*}
\centering\includegraphics[scale=.7]{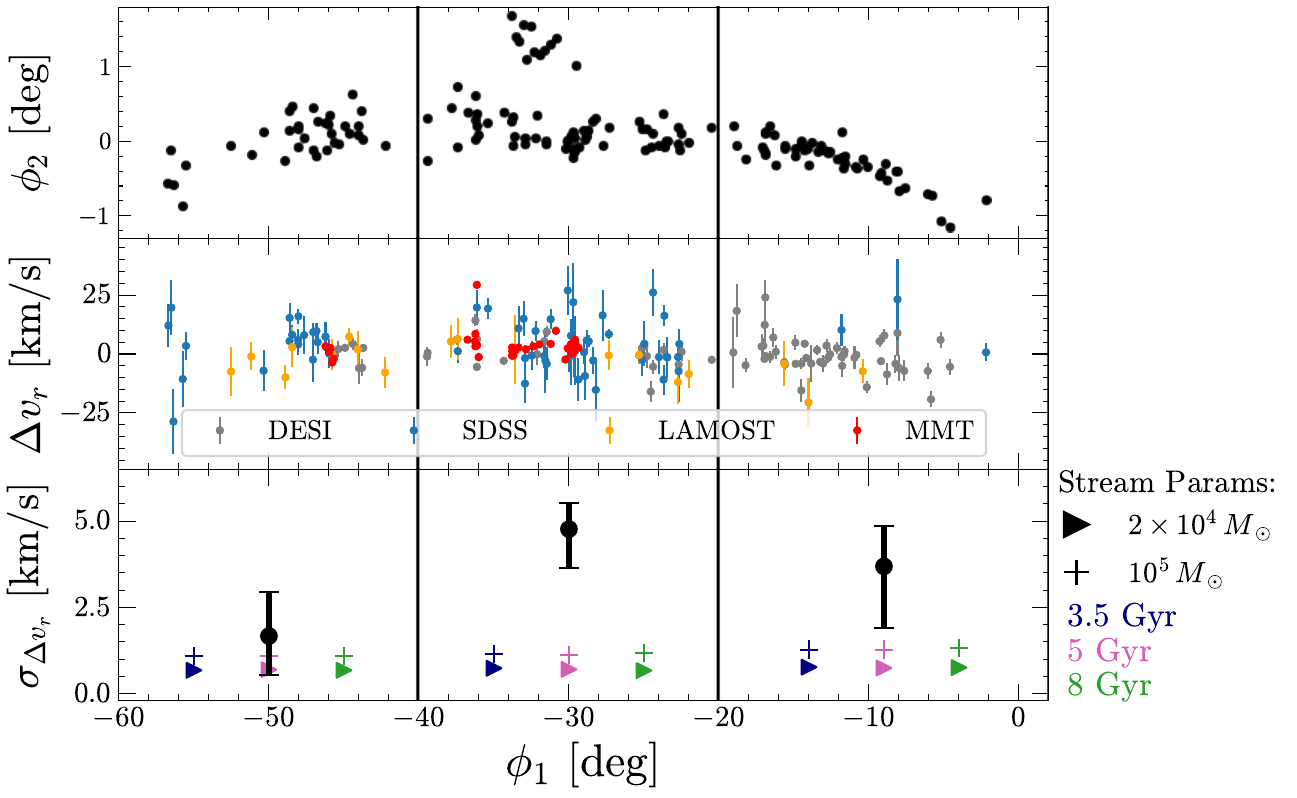}
    \caption{Top panel: Sky-positions of the GD-1 members with radial velocities used in this work. Black vertical lines indicate the three $\phi_1$ bins. Middle panel: solar-reflex correct radial velocities of GD-1 member stars relative to the unperturbed stream's track. Color-coding indicates the survey, consisting of DESI (gray), SDSS (blue) LAMOST (orange), and MMT (red). Bottom panel: black points indicate the intrinsic dispersion measured from the stream, after accounting for the observational uncertainties and systematic sources of scatter (e.g., due to binarity). The colorful points represent the dispersion from unperturbed models in each bin. Triangles (plus symbols) are for a $2\times10^4~M_\odot$ ($10^5~M_\odot$) progenitor. Navy, pink, and green points are for dynamical ages of $3.5, 5,$ and $8~\rm{Gyr}$, respectively. }
    \label{fig: data}
\end{figure*}

The dataset is shown in Fig.~\ref{fig: data}. The top panel plots the sky-positions of the sample (coordinate frame from \citealt{2010ApJ...712..260K}). We apply a solar-reflex correction to the radial velocities using \texttt{Astropy} v4.0 parameters \citep{astropy:2022}. Reflex corrected radial velocities are shown in the middle panel of Fig.~\ref{fig: data}, relative to the unperturbed model (\S\ref{sec: unpert_streams}): $\Delta v_r \equiv v_{r,i} - v_{r,\rm{unpert}}(\phi_{1_i})$.

Next, we measure the radial velocity dispersion along the stream. We bin the stream into three $\phi_1$ segments ($[-60,-40], [-40,-20], [-20,0]~\rm{deg}$). We then measure the intrinsic dispersion within each bin. Although using narrower bins yields consistent dispersion measurements along the stream, we use three broader bins to ensure a sufficient number of stars per bin, thereby providing higher signal-to-noise constraints on the radial velocity dispersion. We bin the data in order to measure the local velocity dispersion along the stream rather than a global dispersion, since the latter will heavily depend on the underlying density distribution of the stream, while local dispersions are less sensitive to the stream's global density. The three bins contain 35, 74, and 51 stars, respectively. Note that the middle bin contains the spur component of the stream, and the majority of the MMT measurements.

\begin{figure*}
\centering\includegraphics[scale=1.2]{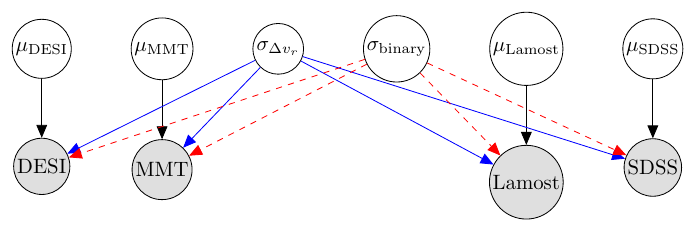}
    \caption{Graphical representation of the statistical model for the radial velocity dispersion. We combine information across four datasets, accounting for the observational errors associated with each measurement. Free parameters include a mean for each dataset ($\mu_X$), radial velocity jitter due to binaries ($\sigma_{\rm binary}$), and the intrinsic radial velocity dispersion ($\sigma_{\Delta v_r}$). }
    \label{fig: DAG}
\end{figure*}

There is considerable scatter in the radial velocities, though errorbars range from a median uncertainty of $0.8~\rm{km/s}$ for the MMT measurements to $6.4~\rm{km/s}$ for SDSS. In order to combine information from surveys with heteroskedastic errors, we utilize a hierarchical Bayesian model to measure the intrinsic radial velocity dispersion of the stream, $\sigma_{\Delta v_r}$. We assume Gaussian distributions for all density functions, and consider three sources of radial velocity scatter: measurements errors ($\sigma_{\rm obs, i}$, reported for each star), scatter in repeat observations across distinct surveys ($\sigma_{\rm binary}$), and the intrinsic scatter of the stream ($\sigma_{\Delta v_r}$). The intrinsic dispersion is a free parameter for each bin, and we also fit the mean $\Delta v_r$ of the four surveys in each bin (i.e., a zero-point correction). The parameter $\sigma_{\rm binary}$ is introduced to account for the possibility of binary stars, which can produce radial velocity jitter over time (e.g., \citealt{Badenes_2018,10.1093/mnras/sty240}). For example, one star in our sample has a nearly $30~\rm{km/s}$ discrepancy across the three surveys (the MMT point near $\phi_1 = -35$ with $\Delta v_r\approx 25~\rm{km/s}$), though this amount of variability is consistent with binarity for the star's $\log{g}$ and surface temperature \citep{Badenes_2018}. This velocity jitter is absorbed into the $\sigma_{\rm binary}$ term, thereby reducing $\sigma_{\Delta v_r}$. We provide a graphical representation of our hierarchical model in Fig~\ref{fig: DAG}, and write out the likelihood function below. 

Let $\Delta v_{r,i}^{(s)}$ represent the radial velocity (relative to the unperturbed stream) for the $i^{\rm th}$ star in survey $s \in \{1,2,3,4\}$, representing \{DESI, MMT, SDSS, LAMOST\}. The mean $\Delta v_r$ for each survey is $\mu_s$. The observational uncertainty reported from each survey is $\sigma_{s,i}$. Survey-to-survey scatter due to (e.g.) binarity is captured by the dispersion parameter $\sigma_{\rm binary}$. The remaining scatter is the intrinsic velocity dispersion, $\sigma_{\Delta v_r}$. Assuming the surveys are independent, the likelihood function for a single star is
   \begin{multline}
    P( \{\Delta v^{(s)}_{r,i}\}_s | \theta) \\= \prod_{s=1}^4\mathcal{N}\left(\Delta v^{(s)}_{r,i} | \mu_{s}, \sigma^2_{s,i} + \sigma^2_{\rm binary} +\sigma^2_{ \Delta v_r}  \right),
\end{multline}
where $\theta$ is the vector of model parameters. If a star is not observed by survey $s$, we take $\sigma_{s,i} \xrightarrow{}10^4~\rm{km/s}$, reflecting the lack of kinematic data for the $i^{\rm th}$ star from survey $s$.
The total likelihood is
\begin{equation}
     \mathcal{L}( \{\Delta v^{(s)}_{r,i}\}_{(s,i)} | \theta) = \prod_{i=1}^N P( \{\Delta v^{(s)}_{r,i}\}_s | \theta),
\end{equation}
and we use Bayes' theorem to derive the posterior $P(\theta | D) \propto \mathcal{L}( D | \theta) P(\theta)$, where $D$ represents the four radial velocity datasets.  We note that our model assumes that the observational errors from each survey are accurate and fully describe the measurement uncertainties. While this is a common approach, a more flexible model could allow for survey-specific error scaling.  However, implementing this would introduce additional complexity and parameters, which is not warranted given the constraining power of the current dataset.

We use a truncated half-normal distribution for priors on $\sigma_{\rm binary}$ and $\sigma_{\rm \Delta v_r}$ with scale parameters of $10~\rm{km/s}$. We also tried a wider prior ($30~\rm{km/s}$) and obtained consistent results. We use central normal priors for $\mu_s$ with a standard deviation of $10~\rm{km/s}$. We implement our model in \texttt{Numpyro} \citep{2019arXiv191211554P} and use the No-U-Turn sampler \citep{10.5555/2627435.2638586}. We find $\sigma_{\rm binary} = 2.6\pm^{1.2}_{1.4}~\rm{km/s}$ at 68\% confidence. Note that this parameter is degenerate with the stream's intrinsic dispersion, $\sigma_{\Delta v_r}$, though we still produce a constraint on $\sigma_{\Delta v_r}$ thanks to the repeat observations of 35 stars in our sample. The 1D marginal posterior on the intrinsic dispersion, $P(\sigma_{\Delta v_r}|D)$, is illustrated in Appendix~\ref{app: intrinsic_dispersion}, Fig.~\ref{fig: sigma_vr_posterior}.

The bottom panel of Fig.~\ref{fig: data} shows our 68\% constraint (black errorbars) on $\sigma_{\Delta v_r}$ across the three bins. Note that the posterior distributions are not Gaussian, so these errorbars are only provided for visual representation, but the full posteriors will be used in our analysis (\S\ref{sec: model}) and are visualized in Appendix~\ref{app: intrinsic_dispersion}. At 68\% we find $\sigma_{\Delta v_r} = 1.7\pm {1.2}~\rm{km/s}$ (left bin; $\phi_1 \in [-60,-40]~\rm{deg}$), $4.8\pm_{1.2}^{0.7}~\rm{km/s}$ (middle bin; $\phi_1 \in [-40,-20]~\rm{deg}$), $3.7\pm_{1.8}^{1.2}~\rm{km/s}$ (right bin; $\phi_1 \in [-20,0]~\rm{deg}$). These values are similar to estimates from \citet{2025ApJ...980...71V} for the intrinsic dispersion of the thin component of GD-1 (i.e., excluding the cocoon component; \citealt{2019ApJ...881..106M}), though note that we have taken significant measures to model survey-to-survey variance, and therefore possible binary contamination. We also compare our results to \citet{2021ApJ...911L..32G}, who used the same MMT dataset to measure the intrinsic velocity dispersion of GD-1. Without accounting for binaries, they report a lower radial velocity dispersion of $2.3\pm0.3~\mathrm{km\,s}$ for the region corresponding to our middle bin. The discrepancy between our results arises from differences in the adopted velocity cuts: we use $|\Delta v_r| < 30~\mathrm{km /s}$, whereas they impose a more stringent criterion of $|\Delta v_r| < 7~\mathrm{km /s}$ since they do not model the full density structure of the stream in each astrometric dimension. When we apply a tighter velocity cut, we recover their result (see Fig.~\ref{fig: sigma_vr_posterior}). Throughout this work, we adopt the wider radial velocity cut, as the tails of the velocity distribution in our simulations are most constraining for dark matter models. Moreover, all stars in our analysis have high membership probabilities based on photometric and astrometric modeling \citep{2025ApJ...980..253S}. We discuss approaches for more robust membership determination without relying heavily on kinematic cuts in \S\ref{sec: discussion}.

\section{Stream Dynamical Modeling}\label{sec: model}
We model the effects of hundreds of subhalo impacts on the GD-1 stream as a function of subhalo parameters using perturbation theory, implemented in the \texttt{streamsculptor}\footnote{\url{https://github.com/jnibauer/streamsculptor}} package \citep{2025ApJ...983...68N,nibauer_2026_19355442}. We provide a summary below. First, we generate a suite of unperturbed models utilizing the particle-spray model from \citet{2025ApJS..276...32C} for stream generation. We then apply perturbations by sampling from a library of subhalo orbits. For each individual subhalo encounter, we precompute perturbation vectors that characterize the leading-order phase-space response of every particle in the model stream.

To capture the cumulative effect of hundreds of subhalos acting on the stream, we rely on the linearity of our perturbative framework. Rather than computing the complex, covariant interactions of multiple subhalos intersecting the stream at varying times, we compute the linear response of every stream particle to each subhalo impact independently. By neglecting the higher-order interaction terms between distinct subhalo perturbations (a valid simplification given that individual low-mass subhalos typically induce linear changes in the stream's morphology and kinematics) the combined dynamical influence of an entire subhalo population reduces to a superposition. While very massive impacts or slow encounters can induce large fractional energy changes in certain particles, these particles develop large perturbation vectors and scatter away from the main stream as discussed in \citet{2025ApJ...983...68N}. Consequently, they typically fall outside of our velocity cut and are clipped in our analysis. We find that this clipping does not introduce a bias, because the scattered particles would not be considered stream members.

In practice, the final perturbed phase-space position of a particle is calculated as its unperturbed coordinate plus the linear sum of the displacement vectors from all $N$ subhalo encounters. Because these encounters are treated independently at linear order, we can efficiently forward-model the stream's morphology under varying subhalo mass functions and scale-radius relations. We achieve this by algebraically rescaling the precomputed individual perturbation vectors by new mass and scale-radius parameters before summing them, completely bypassing the need to resimulate direct orbit integrations for each new realization. All further modeling details, including the subhalo impact sampling and model validation, are presented in \citet{2025ApJ...983...68N}.

\subsection{Unperturbed Streams}\label{sec: unpert_streams}
For the unperturbed stream we consider a range of models to account for uncertainties in the dynamical age of the tidal tails and then stream's progenitor mass. We consider six total models: 3 dynamical ages for each of the progenitor masses $2\times 10^4~M_\odot$ and $10^5~M_\odot$. The lower bound is set by stellar population modeling of probable GD-1 members \citep{2025ApJ...988...45T}, and is consistent with dynamical models \citep{2010ApJ...712..260K}. The upper bound exceeds dynamical estimates of the progenitor's initial mass. However, it is possible that GD-1 is wider and longer than current observational data suggest, so our upper bound provides a conservative choice for the progenitor mass. We consider the integration times [3.5, 5, 8]~\rm{Gyr}. Progenitor locations can be found in the appendix of \citet{2025ApJ...983...68N}. Our priors enclose reasonable estimates for the (now dissolved) GD-1 progenitor and the dynamical age of the tidal tails \citep{2019MNRAS.485.5929W}. Note that the true progenitor could be less massive than $2\times 10^4~M_\odot$ to match the narrow width of the observed stream, excluding the spur component (e.g., \citealt{2019ApJ...880...38B,2019MNRAS.485.5929W,2021ApJ...911L..32G}). However, lower mass progenitors will only lower the dispersion of the model stream, making our choice conservative, since the data prefer higher dispersions than the unperturbed models. Each model stream has 5000 particles, from which the radial velocity dispersion is measured. We have tested that using a larger particle number ($30000$) does not alter our results, though is very memory inefficient.

For the background potential, we utilize \texttt{MilkyWayPotential2022} from \texttt{Gala} \citep{gala}, which is fit to a recent compilation of Milky Way mass measurements. Note that this is a static potential. It is possible to include time-dependence in our global potential modeling (see \citealt{2025ApJ...983...68N}), though GD-1's retrograde orbit and its substantial distance from the Large Magellanic Cloud motivate our static assumption. Small shifts in the track can occur with moderate time-dependence at GD-1's location (e.g., \citealt{2022MNRAS.516.1685D,2024ApJ...969...55N}), though the impact on velocity dispersion is small. We discuss incorporating time-dependence in \S\ref{sec: assumptions}.

\subsection{Dark Matter Subhalos}\label{sec: model_subhalos}
Dark matter subhalos are modeled as Hernquist profiles \citep{1990ApJ...356..359H}:
\begin{equation}
    \rho(r) = \frac{\rho_s}{\left(\frac{r}{r_s} \right)\left(1 + \frac{r}{r_s} \right)^3}
\end{equation}
which is very similar to the universal dark matter density profile of the  Navarro–Frenk–White profile \citep{1997ApJ...490..493N}, but with finite mass owing to a sharper truncation at large radii. We  model the mass-size relation of Hernquist subhalos in CDM as
\begin{equation}\label{eq: r_s_m}
    r_{s}\left(M\right) =  A_r \times \left(1.05~\rm{kpc}\right) \sqrt{M/10^8 M_\odot } ,
\end{equation}
where $M$ is the total mass of the Hernquist profile, and $A_r \equiv r_s / r_{s,\rm{cdm}} =  1$ for Hernquist subhalos in CDM, obtained by fitting the $M_{\rm tidal}-v_{\rm max}$ relation for subhalos in the Via Lactea II simulations \citep{2008Natur.454..735D, 2016MNRAS.463..102E}. We consider subhalos in the range $M\in[10^5,10^9]~M_\odot$. Below $10^5~M_\odot$, subhalos do not produce appreciable heating in a GD-1 like stream \citep{2025ApJ...983...68N}, and impacts above $10^9~M_\odot$ are not expected for GD-1 \citep{2019ApJ...880...38B}. We compute the concentration of Hernquist subhalos using $c_{-2} \equiv R_{200} / r_{-2}$, where $R_{200}$ is the radius within which the subhalo mean enclosed density is 200 times the critical density of the universe, and $r_{-2}$ is the radius at which the logarithmic density slope of the profile is $-2$. For the NFW profile, $r_{-2}$ is the scale-radius. For the Hernquist, $r_{-2} = r_s/2$. We use {\it Planck} 2018 cosmological parameters \citep{2020A&A...641A...6P}.

\begin{figure}
\centering\includegraphics[scale=.58]{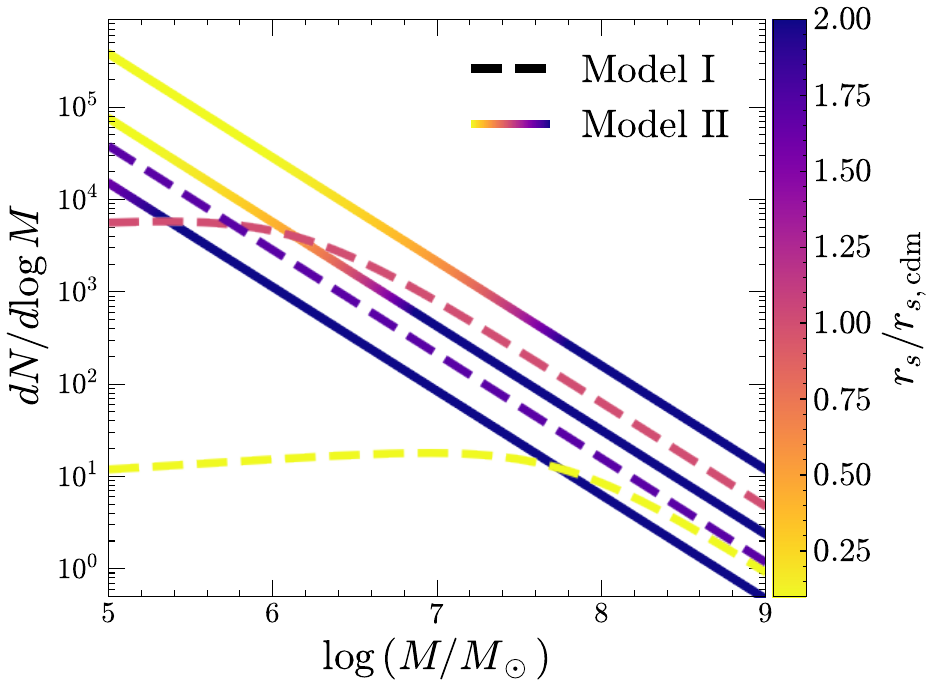}
    \caption{Illustration of the two modeling scenarios. Lines show example draws from the two models, color-coded by the concentration factor $r_s/r_{s,\rm cdm}$. Model I (dashed lines) allows for a suppression in the subhalo mass function at low masses. The concentration factor is independent of mass in this model. Model II allows for a mass-dependent concentration factor, and has a fixed slope corresponding to CDM expectations. Both models allow the normalization of the mass function to vary.  }  
    \label{fig: model_explainer}
\end{figure}

For generating subhalo impacts, we utilize the Einasto functional form for the radial number density of subhalos \citep{2016MNRAS.463..102E,2025ApJ...983...68N}. We use the Via Lactea II subhalo mass function: $dN/dM \propto M^{-1.9}$ \citep{2008Natur.454..735D}. We consider two modeling scenarios to explore the degeneracy between the number of subhalos and their mass-concentration relation. These are summarized in Fig.~\ref{fig: model_explainer}. In summary, Model I allows for a suppression in the subhalo mass function, which is expected under the warm dark matter (WDM) scenario. Model I treats $A_r$ as a constant, independent of subhalo mass. Model II allows $A_r$ to vary as a function of subhalo mass, but does not have a suppression in the SHMF. Both models allow for the normalization of the SHMF to vary. We expand on the two models below.

\textbf{Model I ($\mathbf{r_s/{r_{s,\rm{\textbf{cdm}}}} = \rm{\textbf{constant}}}$):} This model characterizes the warm dark matter (WDM) scenario, assuming the dark matter particle is a thermal relic. The SHMF ($dN/dM$) in this scenario applies a modification to the CDM SHMF as follows:
\begin{equation}
    \left(\frac{dN}{dM} \right)_{\rm WDM} = \left(1 + \gamma\frac{M_{\rm hm}}{M} \right)^{-\beta} \left( \frac{dN}{dM}\right)_{\rm CDM},
\end{equation}
where $\gamma = 2.7$ and $\beta = 0.99$, determined from WDM simulations of a Milky Way mass galaxy based on the Aquarius simulations \citep{2014MNRAS.439..300L}. The half-mode mass, $M_{hm}$, parametrizes a turnover in the subhalo mass function, implying a scarcity of subhalos below $M_{hm}$ compared to CDM expectations ($M_{hm} = 0)$.
The half-mode mass is connected to the WDM particle mass ($m_{\rm WDM}$). For reference, for $M_{\rm hm} = [10^6,10^7,10^8]~M_\odot$, $m_{\rm WDM} = [19,10,5]~\rm{keV}$ \citep{2021ApJ...917....7N}. 
We also define the parameter $f_{\rm sub}$, which is the fraction of the Milky Way's mass in subhalos:
\begin{equation}\label{eq: f_sub}
    f_{\rm sub} = \frac{1}{M_{\rm virial}} \int\limits_{10^5M_\odot}^{10^9M_\odot} M \frac{dN}{dM} dM,
\end{equation}
where we use $M_{\rm virial} = 1.3\times10^{12} M_\odot$ \citep{2017MNRAS.465...76M}. Free parameters in our modeling include the normalization of the SHMF, $M_{hm}$, and the concentration factor $A_r$, which controls the mass-size relation in Eq.~\ref{eq: r_s_m}. Our prior on the normalization of the SHMF is loguniform from effectively 0 to $3\times$ the expected CDM normalization \citep{2008MNRAS.391.1685S,2016MNRAS.463..102E}. We have tested a wider prior, up to 10$\times$ the CDM normalization, and found that more abundant subhalo models can still fit the data if the subhalos are sufficiently diffuse. In other words, from kinematic data alone we find that one can always increase the number of subhalos impacts (the stream impact rate) while making the subhalos less compact to achieve the same stream velocity dispersion. We refer to this as a rate-concentration degeneracy since the impact rate of subhalos will be covariant with the subhalo concentration. In this work we do not explore above $3\times$ CDM in subhalo number density, corresponding to $f_{\rm sub}$ of $\approx 30\%$, significantly higher than CDM expectations ($\approx 8\%$; \citealt{2008MNRAS.391.1685S}). In future work, combining our analysis with density information can likely break the rate-concentration degeneracy. 

For the concentration parameter $A_r$, we adopt a loguniform prior, from $A_r = 0.1$ to $A_r = 2$. For a $10^6~M_\odot$ subhalo these bounds correspond to a concentration ($c_{-2}$) of roughly 400 and 18, respectively. Our prior on $M_{hm}$ is also loguniform from $M_{hm} = 0$ (effectively) to $M_{hm} = 10^{10}~M_\odot$.

\textbf{Model II ($\mathbf{r_s/{r_{s,\rm{\textbf{cdm}}}} = f(M)}, \ \mathbf{M_{\textbf{\rm hm}} = 0}$):} We also consider a separate scenario, where $M_{\rm hm} = 0$ (i.e., the CDM SHMF), and  the parameter $A_r$ in Eq.~\ref{eq: r_s_m} is modeled as a broken power-law in $\log\left(M / M_\odot \right)$. In log-space, the broken power-law is piecewise linear with an offset ($a_1$), two-slopes ($\alpha_1, \alpha_2$) and a breakpoint ($a_2$). The functional form is:
\begin{multline}\label{eq: powerlaw}
    \log\left(\frac{r_s(M)}{r_{s,\rm{cdm}}(M)}\right) \\=
    \begin{cases}
        a_1 + \alpha_1 \left[ \log\left(\frac{M}{M_\odot}\right) - a_2 \right] & \text{if } \log(\frac{M}{M_\odot}) < a_2 \\
        a_1 + \alpha_2 \left[ \log\left(\frac{M}{M_\odot}\right) - a_2 \right] & \text{if } \log(\frac{M}{M_\odot}) \geq a_2.
    \end{cases}
\end{multline}
We require that  Eq.~\ref{eq: powerlaw} is either flat (i.e., $\alpha_1 = \alpha_2 = 0$) or monotonically increasing with mass, allowing us to test whether there is support for models that are more concentrated than CDM expectations, particularly at lower subhalo masses. We also require $\alpha_1 \geq \alpha_2$ to test whether there is support for more concentrated, low-mass subhalos. This is motivated by dark matter particle theories that lead to higher concentrations at lower subhalos masses, such as self-interacting dark matter (SIDM; \citealt{2000PhRvL..84.3760S}) or atomic dark matter \citep{2010JCAP...05..021K}. Note that in the standard cold SIDM scenario, the free-streaming length is negligible, so it is appropriate to set $M_{\rm hm} = 0$. However, there can still be a suppression of SIDM subhalos relative to CDM subhalos due to the enhanced disruption of cored profiles \citep{2025ApJ...991...69N}. For any choice of slope and breakpoint, we limit $r_s$ between $0.1\times $ and $2\times$ the CDM value (same bounds as Model I) so that our linear perturbation theory remains valid. Priors, subject to the above constraints, are uniform in the intervals $a_1 \in [\log(0.1), \log(2)]$, $\alpha_1, \alpha _ 2 \in [0,1]$, $a_2 \in [5,9]$ (the subhalo mass range considered). These choices provide a diverse range of subhalo mass-size relations, and include the CDM mass-size relation when $a_1 = \alpha_1 = \alpha_2 =  0$.

\subsection{Inference}

In this section we describe how the model is connected to the data, and our inference procedure. Even with a fixed subhalo mass function, generating multiple realizations leads to a wide range of outcomes for the stream's velocity dispersion. This variability arises from two main sources of stochasticity. First, the number of subhalo impacts follows a Poisson distribution, so each realization contains a different number of subhalos. Second, in each realization, the subhalos follow different orbits, resulting in varied impact geometries. Together, these factors necessitate running many simulations in order to accurately capture the statistical properties of stream heating caused by subhalo fly-bys.

\begin{figure}
\centering\includegraphics[scale=.65]{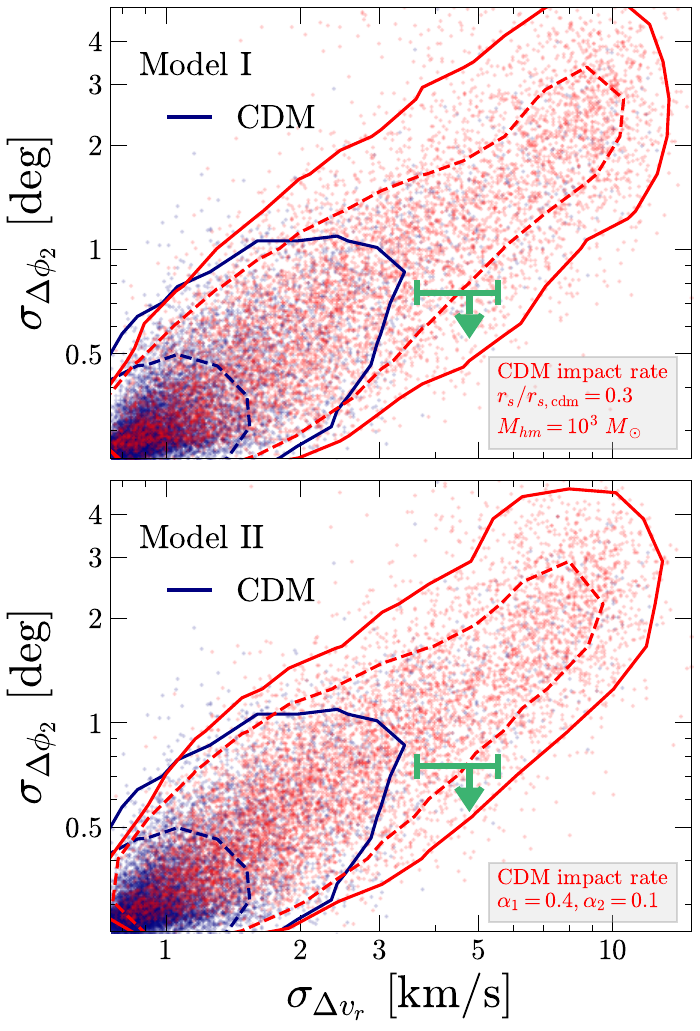}
    \caption{ Illustration of how we connect our model to the data. In both panels we plot the intrinsic velocity dispersion of GD-1 ($\sigma_{\Delta v_r}$, green errorbar), and the upper bound on the stream's width ($\sigma_{\Delta \phi_2}$). The blue points and contour represent many realizations of the model stream under CDM assumptions for the number and size of subhalos. Dashed (solid) contours enclose 68\% (95\%) of samples, and the data sits just outside of the 95\% region. Red points and contours indicate acceptable models that overlap with the data. Model I (top panel) has subhalos with 30\% the scale radius of CDM subhalos, and $M_{hm} = 10^3~M_\odot$. Model II (bottom panel) illustrates the power-law model for subhalo scale-radii, where low-mass subhalos are more concentrated than high mass subhalos. In the $(\sigma_{\Delta v_r}, \sigma_{\Delta \phi_2})$ plane the models are indistinguishable.   }  
    \label{fig: model_data_compare}
\end{figure}

An illustration of how we connect the data to models is provided in Fig.~\ref{fig: model_data_compare} (top panel: Model I, bottom panel: Model II). For both models, we generate $6 \times 10^6$ total simulations, $10^6$ for each combination of progenitor mass and dynamical age. We plot our measurement of the stream's intrinsic dispersion (middle bin; $\phi_1 \in [-40,-20]~\rm{deg}$) in green (68\% confidence level), and the y-axis location is the upper bound on the stream's width ($0.75~\rm{deg}$) from observational data \citep{2025ApJ...988...45T}. We conservatively adopt an upper limit for the stream width (0.75~\rm{deg}, rather than restricting to a narrow range of values, since reported stream widths in the literature vary depending on the criteria used to select stream members. The upper limit we adopt is larger than the thin component of the stream in the literature (e.g., \citealt{2018MNRAS.477.1893D}), though narrower than the cocoon component (e.g., \citealt{2019ApJ...881..106M,2025ApJ...980...71V}), which is not present in the membership modeling from \citet{2025ApJ...980..253S}.

The blue points and containment regions (dashed for 68\% and solid for 95\%) in Model I and II are the same, illustrating, CDM expectations for the distributions of stream widths and velocity dispersion. The data sit outside of the 95\% region for CDM expectations. The red points and contour in the top panel show samples from Model I, assuming the CDM impact rate (i.e., the CDM normalization for the SHMF), $r_s/r_{s,\rm{cdm}} = 0.3$, and $M_{hm} = 10^3~M_\odot$. That is, this model has subhalos that are 70\% more compact than CDM expectations. This leads to an extended tail at higher velocity dispersion and widths, stemming from larger perturbations due to more concentrated subhalos. The velocity dispersion and width of GD-1 are consistent with this model. Note that stream-width is most sensitive to progenitor mass in our simulations. A lower minimum progenitor mass shifts the contours to smaller widths, though our choice of the minimum progenitor mass is conservative in setting the baseline velocity dispersion of the stream, and consistent with the stream's stellar mass \citep{2010ApJ...712..260K,2025NewAR.10001713B}.

In the bottom panel of Fig.~\ref{fig: model_data_compare}, the red points and contour indicate samples from Model II with $\alpha_1 = 0.4, \ \alpha_2 = 0.1$, and $a_1 = -0.6$, $a_2 = 7.8$. The parameters are selected to demonstrate that Model II can produce a distribution of stream widths and velocity dispersions that is nearly identical to that of Model I. For this model, a $10^9~M_\odot$ subhalo has $r_s / r_{s,\rm{cdm}} = 0.4$, while a $10^7~M_\odot$ subhalo has $r_s / r_{s,\rm{cdm}} = 0.1$. This model produces an equally valid solution, and from the $(\sigma_{\Delta v_r}, \sigma_{\Delta \phi_2})$ plane alone cannot be distinguished from Model I.

To determine which samples are accepted or rejected, we compare the model's velocity dispersion to the inferred $\sigma_{\Delta v_r}$ from the data using Approximate Bayesian Computation (ABC; \citealt{10.1214/aos/1176346785}). For a review of ABC methods, see \citealt{10.1093/sysbio/syw077}. In standard ABC, one defines a distance function, $\rho(\rm{data}, \rm{model})$, that compares the model to the data. Samples are generated from the prior, and are accepted only if the distance function is less than a tolerance $\epsilon$. For diminishing $\epsilon$ the ABC posterior converges to the true posterior distribution. We have also experimented with using an explicit likelihood function based on kernel density estimation of the millions of generated samples. However, this approach is highly sensitive to the kernel bandwidth and remains noisy due to the stochastic nature of the problem. In contrast, ABC is better suited for this context, as it efficiently handles stochastic simulations on a sample-by-sample basis and reduces the number of hyperparameters, unlike standard likelihood-based methods that require generating, e.g., millions of realizations per parameter choice.

Our ABC procedure is outlined below. Let $P(\sigma_{\Delta v_r, i} | D)$ represent the posterior distribution for the intrinsic velocity dispersion of the stream in a single bin $i$ (from \S\ref{sec: data}), where $D$ is the velocity measurements from the four datasets.  Let $\theta_{\rm DM}$ represent dark matter parameters (i.e., the normalization of the SHMF, $M_{hm}$, etc.). The posterior probability distribution function (pdf) over the stream's intrinsic dispersion, $\sigma_{\Delta v_r}$, and the dark matter parameters, $\theta_{\rm DM}$ is 
\begin{multline}
    P\left(\sigma_{\Delta v_r}, \theta_{\rm DM} | D\right) \\= \frac{1}{\mathcal{Z}}P\left(D | \sigma_{\Delta v_r} \right) P\left(\sigma_{\Delta v_r} | \theta_{\rm DM}\right) P\left(\theta_{\rm DM}\right),
\end{multline}
where $\mathcal{Z}$ is a normalization constant and we have suppressed the bin index $i$ for simplicity.
The posterior pdf of dark matter parameters given the data is then
\begin{small}
\begin{equation}\label{eq: post}
\begin{split}
    P\left(\theta_{\rm DM} | D\right) &\propto P\left(\theta_{\rm DM}\right)\int d\sigma_{\Delta v_r} P\left(D|\sigma_{\Delta v_r} \right) P\left(\sigma_{\Delta v_r} | \theta_{\rm DM} \right)  \\
    &\propto P\left(\theta_{\rm DM}\right) P\left(D|\theta_{\rm DM}\right),
\end{split}
\end{equation}
\end{small}
where in the last line we defined the likelihood, $P\left(D|\theta_{\rm DM}\right)$, which is equal to the integral in Eq.~\ref{eq: post}. 
In our framework, the term $P\left(\sigma_{\Delta v_r} | \theta_{\rm DM}\right)$ --- the likelihood component representing the distribution of stream velocity dispersion values in a spatial bin given a dark matter model --- will be produced through simulation. 
We now introduce the ABC posterior, which utilizes simulations to approximate the true posterior, Eq.~\ref{eq: post}. Let $D_{\rm sim}$ represent a realization of a simulated stream with parameters $\theta_{\rm DM}$. The joint ABC posterior is
\begin{multline}
    P_{\rm ABC}\left(\theta_{\rm DM}, D_{\rm sim} | D \right) \\ \propto P\left(\theta_{\rm DM} \right)P\left(D|\theta_{\rm DM}\right)K_{\epsilon}\left(D, D_{\rm sim}\right),
\end{multline}
where we have introduced the ABC kernel $K_\epsilon$, with tolerance parameter $\epsilon$. We will take $K_\epsilon$ to be an indicator function, $\mathbb{I}\{\rho\left(D,D_{\rm sim}\right)< \epsilon \}$, which is $1$ when the argument is satisfied and $0$ otherwise, and $\rho(a,b)$ is a Euclidean distance measure between $a$ and $b$. The target ABC posterior is
\begin{multline}\label{eq: abc_post}
   P_{\rm ABC}\left(\theta_{\rm DM}, | D \right) \\\propto P\left(\theta_{\rm DM}\right)\int dD_{\rm sim} P\left(D|\theta_{\rm DM}\right)\mathbb{I}\{\rho\left(D,D_{\rm sim}\right)< \epsilon \}.
\end{multline}
From Eq.~\ref{eq: abc_post}, as $\epsilon \xrightarrow{}0$ the ABC posterior converges to the true posterior.

For our distance measure we compute the Euclidean distance between the simulated velocity dispersion, $\sigma_{\Delta v_r,\rm{sim}}$, and a realization of the intrinsic velocity dispersion $\sigma_{\Delta v_r} \sim P\left(\sigma_{\Delta v_r} | D\right)$ from \S\ref{sec: data}. The sampling procedure that follows from Eq.~\ref{eq: abc_post} is summarized below:
\begin{enumerate}
    \item Sample $\theta_{\rm DM}$ from the prior, $\theta_{\rm DM} \sim P(\theta_{\rm DM})$.
    
    \item Simulate a stream realization under $\theta_{\rm DM}$. The intrinsic velocity dispersion is $\sigma_{\Delta v_r, \rm{sim}}$. 
    
    \item Sample $\sigma_{\Delta v_r} \sim P(\sigma_{\Delta v_r} | D)$, representing a realization of the test statistic consistent with the data. 

    \item If $\rho\left(\sigma_{\Delta v_r, \rm{sim}},  \sigma_{\Delta v_r} \right) < \epsilon$, accept the sample. Otherwise, reject.

    \item Repeat steps 1--4 many times.
\end{enumerate}

The model streams are binned in the same way as the data; however, to eliminate any phase dependence in our constraint, we permute the ordering of the model bins and select the permutation that minimizes the distance to the data (i.e., for three bins we test all six possible orderings). This ensures that a model is not penalized if it reproduces the correct dispersion in one part of the stream, but shifted in phase relative to our binning choice. We have validated that our results are unchanged if we do not shuffle bins, but we retain more samples by eliminating phase-dependence and require fewer overall simulations.

 For diminishing $\epsilon$, ABC converges to the true underlying posterior, and has been used in prior studies of stellar streams and dark matter subhalo populations \citep{2017MNRAS.466..628B,2021JCAP...10..043B}. We choose $\epsilon$ such that 0.05\% of samples are accepted (corresponding to $\epsilon = 0.3$). This is the smallest acceptance fraction that generates visually smooth contours in parameter space. For small deviations below this $\epsilon$ there is no appreciable change in our constraints. Even above this $\epsilon$ our results are converged, until $\approx 1\%$ of samples are accepted. We have validated that this routine recovers the true $\theta_{\rm DM}$ when applying our analysis to simulated streams with known ground truths. We also reject samples where the stream width exceeds $0.75~\rm{deg}$, based on the maximum width from \citet{2025ApJ...988...45T}. In some simulations, the stream is completely destroyed. To filter these out, we require that the number of stars in each bin, relative to the maximum number of stars in any bin, is greater than 0.3. This is conservative, since the ratio for the actual dataset is $\approx  0.5$. 

\section{Results}\label{sec: results}

\subsection{Unperturbed Models}
We first comment on the velocity dispersion of unperturbed models, generated without any subhalos. The dispersion for our six models in each bin is visualized in the bottom panel of Fig.~\ref{fig: data}, shown as colorful symbols. Triangles and $+$ symbols are for a progenitor mass of $2\times 10^4~M_\odot$ and $10^5~M_\odot$, respectively. Navy, pink, and green points correspond to $3.5, 5,$ and $8~\rm{Gyr}$, respectively. We add offsets in $\phi_1$ for each point in the bins so that they do not overlap, though this is only for visual comparison. First we note that the unperturbed models have approximately constant velocity dispersions across the stream. The velocity dispersion is mostly sensitive to the progenitor mass, and less sensitive to the stream's dynamical age. The median velocity dispersion for the lower (higher) mass model is 0.7~\rm{km/s} (1.1~\rm{km/s}). For the highest dispersion bin that we measure from the data (the middle bin; $\phi_1 \in [-40,-20]~\rm{deg}$), the unperturbed model velocity dispersions differ from the observations by $4.1\sigma$ for the lower-mass progenitor with an age of $5~\rm{Gyr}$, and $3.7\sigma$ for the higher-mass progenitor of the same age (where $\sigma$ is the Gaussian uncertainty of the measured intrinsic dispersion). The unperturbed velocity dispersions in the left bin ($\phi_1 \in [-60,-40]~\rm{deg}$) are consistent with the data, while the data prefer higher values than the models in the right bin ($\phi_1\in[-20,0]~\rm{deg}$) at the $2\sigma$ level. We have validated that the velocity dispersions from the unperturbed stream models match direct $N-$body simulations in Appendix~\ref{app: nbody}.

\subsection{Perturbed Models}
We now explore constraints on the number and size of dark matter subhalos from the measured velocity dispersion of GD-1. 

\begin{figure}
\centering\includegraphics[scale=.4]{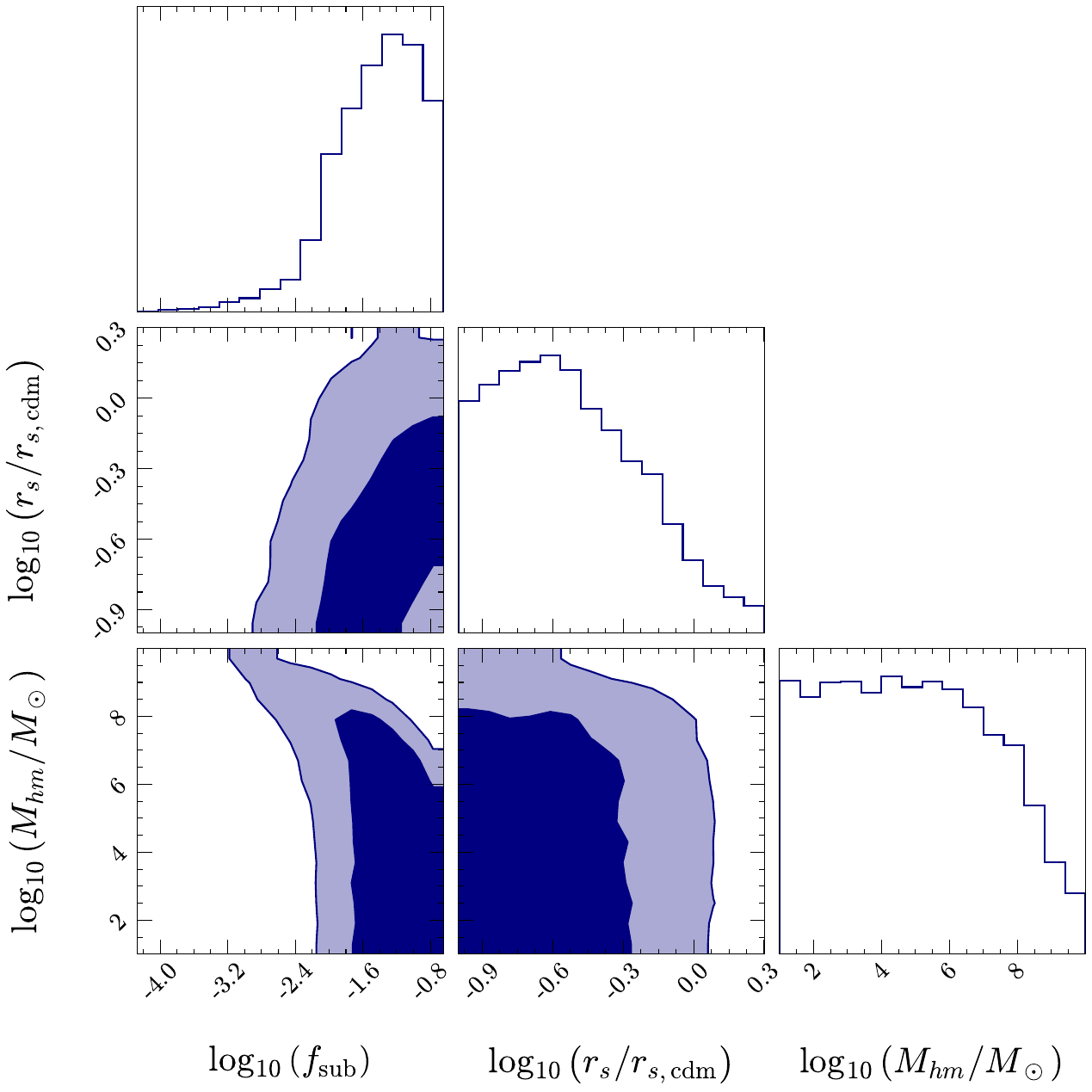}
    \caption{Posterior distribution for Model I. Dark and light blue indicate regions of 68 and 95\% confidence, respectively. There is a degeneracy between the number of subhalos ($f_{\rm sub}$) and the dark matter half-mode mass ($M_{hm}$), such that higher $M_{hm}$ implies fewer subhalos. Fewer subhalos imply more compact central densities (lower $r_s/r_{s,\rm{cdm}}$).   }
    \label{fig: corner_modelI}
\end{figure}

\begin{figure*}
\centering\includegraphics[scale=.62]{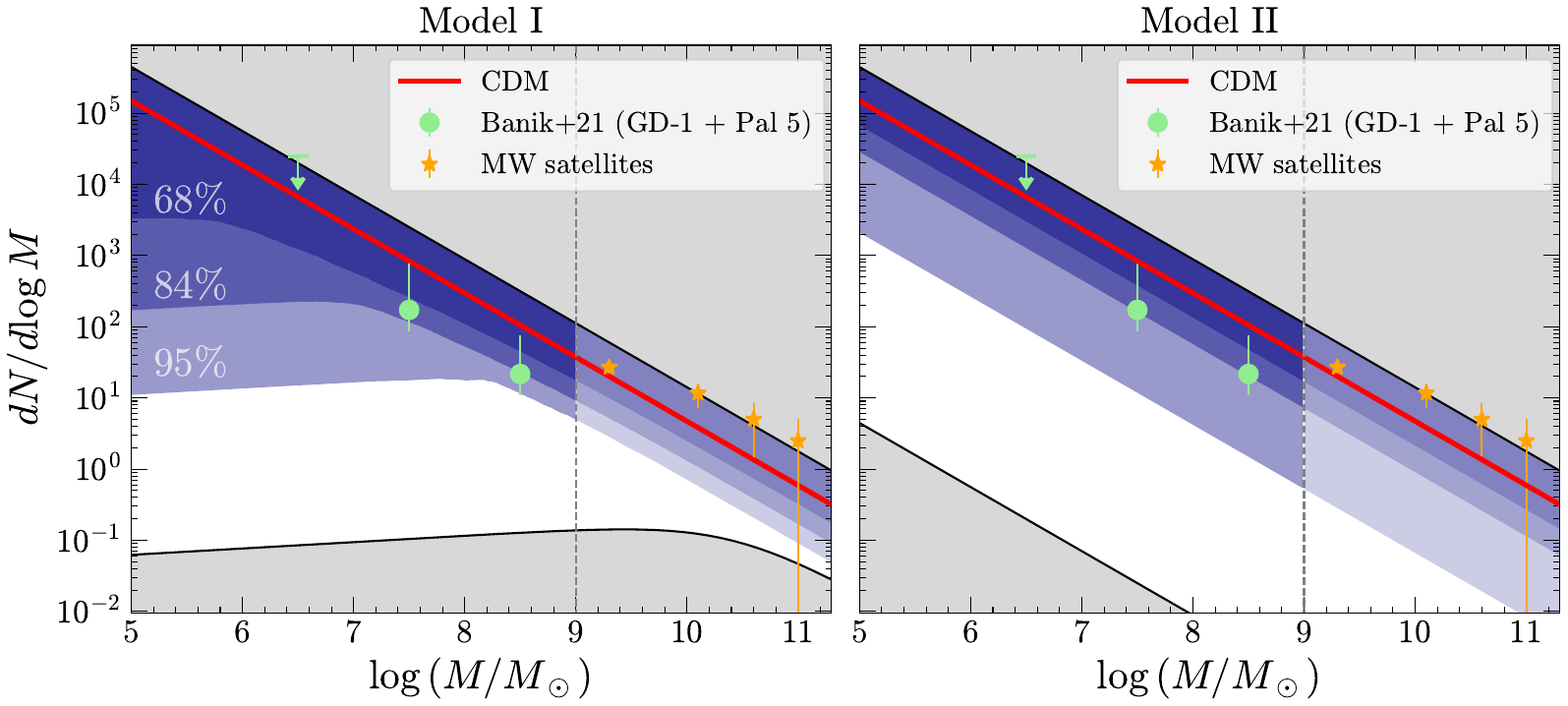}
    \caption{Constraints on the subhalo mass function for Model I (left) and Model II (right). Dark, medium, and light blue indicate regions of 68, 84, and 95\% confidence, respectively. Both panels contain the CDM mass function in red. Black lines indicate prior bounds, gray regions are not sampled by our model. Model I allows for a non-zero half-mode mass, and a suppression of low-mass subhalos. Model II has no suppression in low-mass subhalos. Green points indicate constraints from the GD-1 and Pal 5 streams \citep{2021JCAP...10..043B}, orange points are from number counts of Milky Way satellites \citep{2020ApJ...893...48N}. }
    \label{fig: shmf}
\end{figure*}

\begin{figure}[htp!]
\centering\includegraphics[scale=.7]{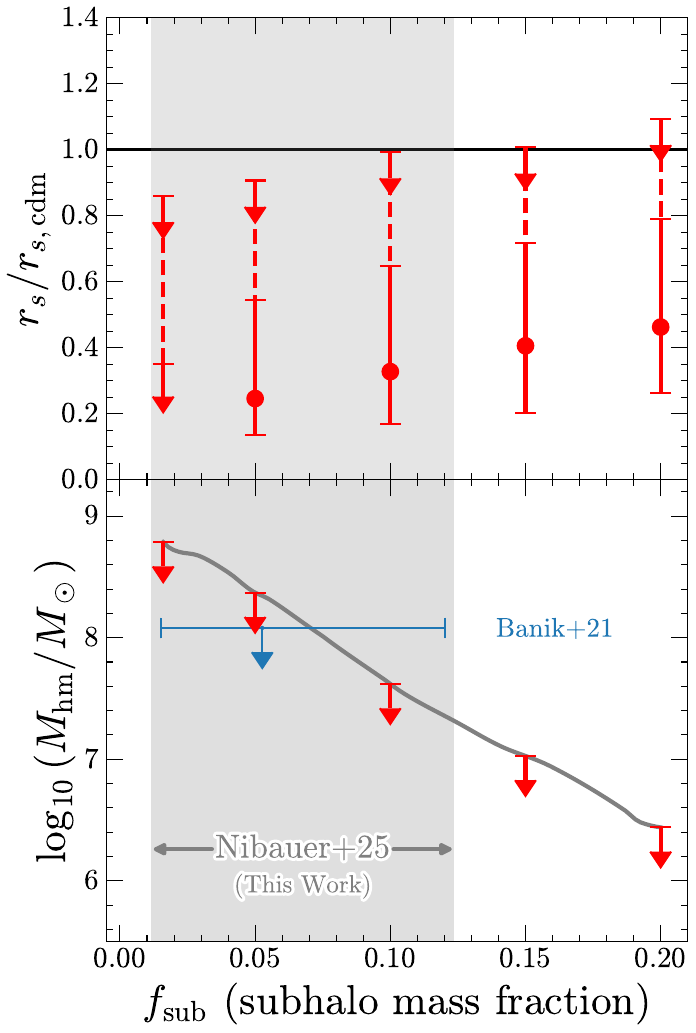}
    \caption{Constraints on the subhalo mass function and mass-size relation for Model I. The gray band indicates our constraint on the fraction of mass in subhalos (68\%). Top panel: constraints on the scale radius of subhalos relative to CDM expectations. For this model we assume $r_s/r_{s,\rm{cdm}}$ is a constant, independent of subhalo mass. Errorbars represent 68\% intervals, while the dashed red line is the 95\% upper limit. Bottom panel: the gray curve and red arrows indicate the 95\% upper limit on $M_{hm}$ as a function of the subhalo mass fraction. Constraints from \citet{2021JCAP...10..043B} are overplotted in blue. At higher mass fractions we require lower values of $M_{hm}$, consistent with more subhalos at lower masses. }
    \label{fig: sliced_constraints}
\end{figure} 

\begin{figure}
\centering\includegraphics[scale=.55]{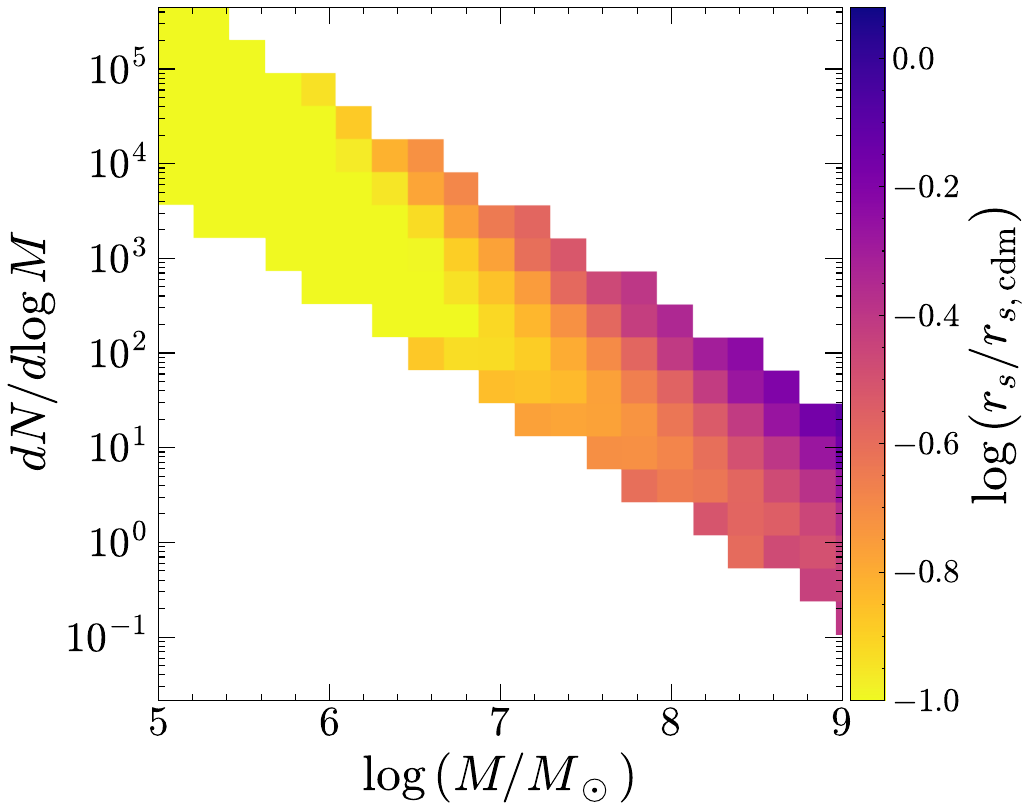}
    \caption{Constraints on the mass-size relation for Model II. Here we plot the 84\% containment interval for the Model II SHMF, color-coded by the median inferred $\log(r_s/r_{s,\rm{cdm}})$. At lower normalizations of the mass function, there is a preference for increasingly more compact, low-mass subhalos.  }
    \label{fig: ModelII_rs_SHMF}
\end{figure}

Constraints on the subhalo mass function and the mass-size relationship for Model I are shown in Fig.~\ref{fig: corner_modelI}. Two important degeneracies are present. First, there is a relationship between the subhalo mass fraction ($f_{\rm sub}$) and the half-mode mass ($M_{hm}$): when $M_{ hm}$ is higher, the fraction of mass in subhalos is lower, because subhalos with masses below $M_{hm}$ are strongly suppressed. Second, there is a relationship between $f_{\rm sub}$ and the concentration factor ($r_s/r_{s,\rm{cdm}}$): if $f_{\rm sub}$ is lower, there are fewer subhalo impacts, which means that the remaining subhalos need to be more compact to account for the observed velocity dispersion. We expect that including stream density information in our modeling will help resolve this degeneracy.

Now we explore constraints on the fraction of mass in subhalos from Model I and Model II. The full Model II posterior is provided in Appendix.~\ref{app: model_II}. Our constraint on the fraction of mass in subhalos is 
\begin{equation}\label{eq: f_sub_constraint}
\begin{split}
    f_{\mathrm{sub}} &= 0.05^{+0.08}_{-0.03} \ \ (\rm{Model \ I}, 68\%) \\
    f_{\mathrm{sub}} &> 0.04 \ \ (\rm{Model \ II}, 68\%).
\end{split}
\end{equation}
These results are consistent with CDM expectations, which predict $f_{\rm sub}\approx 7.5\%$ \citep{2008MNRAS.391.1685S,2023ApJ...945..159N}. Using the density structure of the same stream, \citet{2021MNRAS.502.2364B} inferred a subhalo  mass fraction of $f_{\rm sub} = 0.05 \pm^{0.07}_{0.03}$, consistent with our results. For Model II, we run into our prior at high $f_{\rm sub}$, so we only quote lower limits. The 95\% lower limit for Model I is  $f_{\rm sub} >0.5\%$, corresponding to  $5.9\times 10^9~M_\odot$ of the Milky Way's mass in dark subhalos. The 95\% lower limit for Model II is $f_{\rm sub}>0.01$. Lower values for $f_{\rm sub}$ are preferred for Model I compared to Model II. This is because in Model I, we require all subhalos to have the same $r_s/r_{s,\rm{cdm}}$, and allow for a non-zero half-mode mass. In Model I, it is sufficient to have a single impact with a subhalo that is highly concentrated. In Model II we set $M_{hm} = 0$, so there are subhalos down to $10^5~M_{\odot}$ for each realization.

Constraints on the subhalo mass function for both models are shown in Fig.~\ref{fig: shmf} (Model I in the left panel, Model II in the right panel). The dark, medium, and light blue shaded regions represent 68, 84, and 95\% confidence levels, respectively. Black lines indicate the lower and upper limits of our prior, and gray regions are outside of the prior volume. We extend the inferred mass function to $M>10^9~M_\odot$ (right of the dashed line) in order to compare with constraints from Milky Way satellite galaxies. The CDM mass function is shown in red. Green error bars show measurements of the subhalo mass function based on the density structure of the GD-1 and Pal 5 stellar streams \citep{2021JCAP...10..043B}, while the orange points represent constraints derived from counting the number of Milky Way satellite galaxies \citep{2020ApJ...893...48N}.  The satellite counts from the Milky Way assume a stellar-mass to halo-mass relation \citep{2020ApJ...893...48N}.

Both models are consistent with the number of subhalos predicted by CDM, and also allow for $3\times$ the CDM normalization (though more numerous subhalos imply lower concentration factors; Fig.~\ref{fig: corner_modelI}). The 68\% lower limit on the SHMF normalization is $0.5\times$ the CDM normalization for Models I and II. This limit is in agreement with baryonic disruption of subhalos, which can lead to a factor of $\approx 2-3$ reduction in the number of subhalos within $50~\rm{kpc}$ (e.g., \citealt{2016MNRAS.458.1559Z}). Constraints from both models are also in agreement with those derived from the density structure of GD-1 and Pal 5 \citep{2021JCAP...10..043B}, though there is a slight preference for more numerous subhalos between $10^7$ and $10^9~M_\odot$. When extrapolating our constraints above $10^9~M_\odot$, our results are in agreement with number counts of the Milky Way satellites. 

Model I, which allows for a suppression in the SHMF through the half-mode mass $M_{ hm}$, naturally prefers fewer total subhalos compared to Model II. The parameter degeneracies in Fig.~\ref{fig: corner_modelI} illustrate this mechanism: when $M_{hm}$ is high, indicating a strong suppression of low-mass subhalos, the scarcity of these subhalos drives the overall subhalo mass fraction $f_{\rm sub}$ to lower values. To compensate for this deficit of subhalos and still reproduce the stream's elevated velocity dispersion, Fig.~\ref{fig: corner_modelI} also shows that the remaining subhalos must be significantly more compact as $f_{\rm sub}$ decreases. To physically quantify this limit, we compute the expected subhalo impact rate for a highly compact scenario supported by the data, specifically setting $r_s/r_{s,\rm{cdm}} = 0.1$. This calculation reveals that a single close encounter, defined as an impact within two subhalo scale radii ($2r_s$) from a compact subhalo, is entirely sufficient to generate the observed velocity dispersion.

We now explore constraints on the mass-size relation from Model I in Fig.~\ref{fig: sliced_constraints}.
The 68\% constraint on $f_{\rm sub}$ is shaded in gray. The constraint from \citet{2021JCAP...10..043B} is shown as the blue errorbar. In the top panel of Fig.~\ref{fig: sliced_constraints}, we plot our constraints on the mass-size relation of the subhalos relative to CDM expectations, $r_s/r_{s,\rm{cdm}}$. Red errorbars represent 68\% intervals as a function of $f_{\rm sub}$, and downwards arrows are 95\% upper limits. For lower mass fractions our constraint prefers increasingly more compact subhalos. At the CDM predicted mass fraction of $f_{\rm sub} \approx 7.5\%$, we find a preference for subhalos more compact than CDM, with $r_s/r_{s,\rm cdm} = 0.3 \pm^{0.3}_{0.1}$ at the 68\% level. At the 95\% level our constraint on $r_s/r_{s,\rm{cdm}}$ is consistent with 1 for $f_{\rm sub} > 0.1$, though prefers slightly more compact values, $r_s/r_{s,\rm{cdm}} \lesssim 0.9$, for $f_{\rm sub} < 0.1$. Note that the 68\% constraint for $f_{\rm sub} = 0.01$ is an upper limit, with the model preferring $r_s/r_{s,\rm{cdm}} < 0.4$.  The upper limit on $r_s/r_{s\rm{cdm}}$, marginalized over all Model I parameters is $r_s/r_{s,\rm{cdm}} < 1$ at 95\% confidence. Upper limits are dominated by our uncertainty in the intrinsic velocity dispersion of the stream, and a larger number of radial velocity members with precise $v_r$ measurements will improve this limit.

In the bottom panel of Fig.~\ref{fig: sliced_constraints} we plot the 95\% upper limits on the dark matter half-mode mass, $M_{hm}$, as a function of the subhalo mass fraction (gray line and red downwards arrows). There is a clear degeneracy between the subhalo mass fraction and $M_{hm}$: for lower mass fractions, $M_{hm}$ is higher to account for the scarcity of subhalos. At a mass fraction of $f_{\rm sub} = 0.01$, the 95\% upper limit is $M_{\rm hm} < 10^9~M_\odot$. Our simulations go up to $M = 10^9~M_\odot$, so at 95\% confidence we cannot rule out an impact with a massive ($M\lesssim 10^9~M_\odot$) compact ($r_s/r_{s,\rm{cdm}} \lesssim 0.4$) subhalo.  At $f_{\rm sub} = 0.2$, the 95\% upper limit is $\log\left(M_{\rm hm}/M_\odot\right) < 6.4~M_\odot$, the regime of numerous subhalo impacts. At this mass fraction, $r_s/r_{s,\rm{cdm}} = 0.5 \pm^{0.3}_{0.2}$. Note, however, that a mass fraction greater than $12\%$ is disfavored at the 68\% level.

We draw a comparison to \citet{2021JCAP...10..043B}, who also constrains $M_{\rm hm}$ from the GD-1 stream. Their work uses density information without kinematics, whereas here we use only kinematics, providing an independent test of their findings with a new phase-space dimension. However, a direct comparison is difficult because we include the spur in our modeling, while  \citet{2021JCAP...10..043B} did not. \citet{2021JCAP...10..043B} finds a 95\% upper limit on $\log_{10}\left(M_{hm}/ M_\odot \right)$ of $8.1$ at a median mass fraction of $f_{\rm sub} = 0.06$. This is consistent with our result, $\log_{10}\left(M_{hm}/M_\odot\right) < 8.2$ at the same mass fraction. Note, however, that while the constraint on $M_{hm}$ is consistent, our constraint is in slight tension with \citet{2021JCAP...10..043B}, because we require substantially more compact subhalos, whereas they fix $r_s / r_{s, \rm{cdm}} = 1$. It is probable that including the spur component in their analysis will reveal a preference for more compact subhalos at the same impact rate, since we find the bin containing the spur to have the highest intrinsic velocity dispersion. We discuss our preference for more compact subhalos in \S\ref{sec: summary_and_discuss}. 

\begin{figure}[tp!]
\centering\includegraphics[scale=.63]{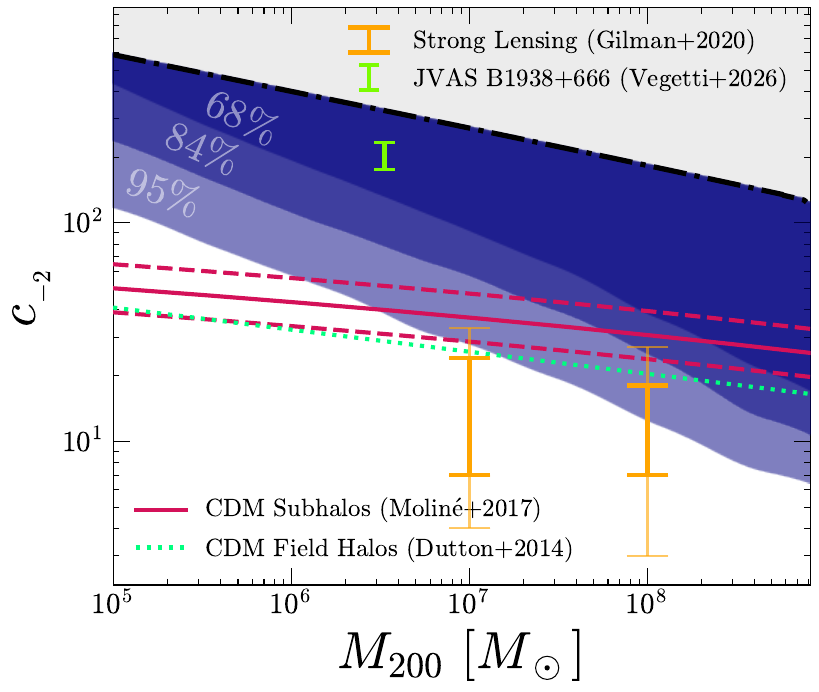}
    \caption{Constraints from Model II on the concentration ($c_{-2}$) of subhalos as a function of $M_{200}$, marginalized over the mass fraction in subhalos. In this model, we treat $r_s/r_{s,\rm{cdm}}$ as a broken power-law in $\log{M}$. Dark, medium, and light blue indicate regions of 68, 84, and 95\% confidence, respectively. The solid red line shows the CDM mass-concentration relation for subhalos \citep{2017MNRAS.466.4974M}, and dashed lines indicate the $1\sigma$ scatter around this relation. The dotted light blue line represents the same relation for field halos \citep{2014MNRAS.441.3359D}. The gray region is not sampled in our model. Constraints from 11 strong gravitational lenses are shown as thick and thin orange errorbars for 68 and 95\% confidence intervals, respectively \citep{2020MNRAS.492L..12G}. The derived concentration of the perturber in the strong lens system JVAS B1938+666 is shown in green at 68\% confidence \citep{2025NatAs...9.1714P,2026NatAs.tmp....5V}.
    }
    \label{fig: concentration}
\end{figure} 

We now consider constraints on the mass-size relation of subhalos under Model II. In this model, we assume a CDM mass function ($M_{hm} = 0$), and model $r_s/r_{s,\rm{cdm}}$ as a function of the subhalo mass, $M$, using a broken power-law in $\log(M)$ (Eq.~\ref{eq: powerlaw}). In Fig.~\ref{fig: ModelII_rs_SHMF} we illustrate the degeneracy between the number and size of subhalos in Model II. The 84\% high confidence region of the subhalo mass function is plotted, color-coded by the posterior median of $\log(r_s/r_{s,\rm{cdm}})$. Brighter colors indicate more compact subhalos. Here we can see that there is a preference for more compact subhalos below $M \lesssim 10^7~M_\odot$, particularly for lower normalizations of the SHMF. This is the same impact rate-concentration degeneracy seen in Fig.~\ref{fig: corner_modelI} for Model I, but for the more complex Model II. The preference for more compact low-mass subhalos can be understood in terms of impact rates. The impact rate for subhalos with $M > 10^8~M_\odot$ is of order a few, and approximately $\mathcal{O}(1)$ impacts within $2r_s$. In contrast, lower-mass subhalos are more numerous, with impact rates of order $\mathcal{O}(20)$ out to $2r_s$. Consequently, lower-mass subhalos frequently encounter the streams in our models. However, at CDM concentrations their contribution to stream heating remains small. At higher concentrations, low-mass subhalos are able to reproduce the intrinsic dispersion of GD-1 in most of our simulations. High-mass subhalos, while rarer, produce strong perturbations even at CDM concentrations, so increasing their concentration is not necessary to explain the GD-1 velocity dispersion. Fig.~\ref{fig: ModelII_rs_SHMF} shows only a posterior median, and there is substantial variance in the compactness parameter that we will highlight below.

We cast our constraints on the mass-size relation from Model II in terms of the subhalo concentration, $c_{-2}$ (see \S\ref{sec: model_subhalos} for concentration definition). Our constraint on $c_{-2}$ as a function of $M_{200}$ is shown in Fig.~\ref{fig: concentration}, marginalized over the subhalo mass fraction. The dot-dashed black curve is the upper bound of our prior ($r_s/r_{s,\rm{cdm}} = 2$), and the grayed out region above this line is not sampled. Dark, medium, and light blue are regions of 68, 84, and 95\% confidence, respectively. The red line indicates the CDM prediction for subhalos, and red dashed lines indicate the expected scatter around this relation \citep{2017MNRAS.466.4974M}. We also include the mass-concentration relation for field halos, which tend to have lower concentrations (dashed teal line; \citealt{2014MNRAS.441.3359D}). Theoretical concentrations are extrapolated below $10^6~M_\odot$ due to numerical resolution in those works. We overplot constraints derived from 11 strong gravitational lenses at 68\% confidence (thick orange errorbars) and 95\% confidence (transparent thin orange errorbars; \citealt{2020MNRAS.492L..12G}). The green errorbar represents the derived concentration from the $\approx 10^6~M_\odot$ object discovered in the strong lens system JVAS B1938+666 using the gravitational imaging technique \citep{2025NatAs...9.1714P}. To derive this, we use the NFW fitted value and errorbar on $c_{\rm{v}}$ from \citet{2026NatAs.tmp....5V} and convert to $c_{-2}$. We also use the characteristic mass reported in \citet{2026NatAs.tmp....5V} to derive $M_{200} \approx 3.3\times10^6~M_\odot$.

At higher subhalo masses, $M_{200}\gtrsim 10^7~M_\odot$, we find agreement with the CDM mass-concentration relation and strong lensing (orange errorbars) at the 95\% confidence level. For lower subhalo masses there is a preference for subhalos with higher concentrations than CDM. For $M_{200}=10^6~M_\odot$ ($M_{200}= 5\times 10^5~M_\odot$) we find $c_{-2} > 190$ ($c_{-2}> 238$) at 68\% confidence. For the same masses, the 95\% lower limit is  $c_{-2} > 57$ ($c_{-2}> 70$). At $M_{200} = 10^5~M_\odot$ we prefer $c_{-2}> 422$ at 68\% confidence, and $c_{-2}> 115$ at 95\% confidence. 

We note that the constraints from strong lensing (orange errorbars) are slightly below the CDM expectations from subhalos and field halos. When adopting WMAP9 cosmological parameters and the mass-concentration relation from \citet{Bullock01}, the CDM expectation shifts downwards to become more consistent with strong lensing (see Fig.~4 of \citealt{2020MNRAS.492L..12G}). Here we use {\it Planck} cosmological parameters, which \citet{Bullock01} is not calibrated against.

\section{Discussion}\label{sec: discussion}

\subsection{Velocity Dispersion Measurement}
We first discuss our measurement of the stream's velocity dispersion. We combine four radial velocity datasets and constrain the contribution of the velocity dispersion due to binarity using repeat observations. Our constraint on the velocity dispersion of the stream in the central region, $\phi_1\in[-40,-20]$, is $\sigma_{\Delta v_r} = 4.8\pm^{0.7}_{1.2}~\rm{km/s}$. This is consistent with recent works utilizing the same datasets and different statistical methodologies \citep{2025ApJ...980...71V, 2025ApJ...988...45T}. However, our constraint is at odds with \citet{2021ApJ...911L..32G}, who finds a lower dispersion using high precision radial velocities from MMT ($\sigma_{\Delta v_r} = 2.3\pm0.3~\rm{km/s}$). The same dataset is included in our work. The discrepancy arises from the velocity cut adopted in each study: we use $|\Delta v_r| < 30~\rm{km/s}$, while they use $|\Delta v_r| < 7~\rm{km/s}$. When we adopt the same velocity cut as \citet{2021ApJ...911L..32G}, our results are consistent. We note, however, that even with the $|\Delta v_r| < 30~\rm{km/s}$, the posterior distribution over $\sigma_{\Delta v_r}$ still has a low dispersion tail (Fig.~\ref{fig: sigma_vr_posterior}). Most importantly, we apply the same velocity cut to both the model and the data, ensuring a consistent comparison. We also note that our choice of the standard deviation for characterizing the stream's kinematics does not require an underlying Gaussian distribution for our analysis to be unbiased. We use the standard deviation as a summary statistic for ABC, which makes no assumptions about the underlying data distribution. More informative statistics can be used in future work when there is additional precise radial velocity data available.

A velocity dispersion of $2.3~\rm{km/s}$ for GD-1 falls within the range predicted by CDM models for the number and concentration of subhalos (see Fig.~\ref{fig: model_data_compare}). We adopt a wider velocity cut since we only select high confidence stream-members based on density modeling of the stream in the other five phase-space dimensions \citep{2025ApJ...980..253S}. Additionally, in our models, the tails of the velocity distribution provide the strongest constraints on subhalo properties. With sufficiently large radial velocity samples, future studies could improve our membership selection by modeling the 6D distribution of stream members to assess the extent of GD-1's velocity distribution. Additionally, binarity can be addressed and further constrained with multi-epoch spectroscopy across a larger sample of GD-1 member stars. Upcoming radial velocity datasets will make this possible (e.g., Via Collaboration, in prep). Further improvements in membership modeling without heavily relying on kinematic cuts can come from chemical abundances (e.g., \citealt{2025ApJ...989L..52Z}). 

\subsection{Modeling Degeneracy}
We have presented constraints on the number and size of dark matter subhalos using two models. The first (Model I) allows for a suppression in the subhalo mass function at low subhalo masses, and assumes a constant mass-size relation ($r_s/r_{s,\rm{cdm}}$) that is the same across all subhalos masses. The second (Model II) has no low-mass suppression in the mass function, but allows for a mass-dependent mass-size relation. We find that both models are capable of describing the GD-1 velocity dispersion equally well (Fig.~\ref{fig: model_data_compare}). In the case of Model I, it is possible to have a suppression in the subhalo mass function if the remaining high-mass subhalos are $\approx 60-70\%$ more compact than CDM expectations. For Model II, there is still a preference for more compact subhalos compared to CDM, though only at the low-mass end where subhalos are more numerous. We expect that modeling stream density with kinematics will break this modeling degeneracy, since a single impact with a very compact halo produces a different density signature than multiple impacts with low-mass, compact subhalos.

In both models, we observe a degeneracy between the number of subhalo impacts and the scale radii of subhalos. Specifically, a higher number of subhalo impacts can produce the same velocity dispersion if the subhalos are more diffuse. This impact rate–concentration degeneracy may be resolved by jointly analyzing the stream's density and kinematics. For example, when modeling GD-1's density we find that density fluctuations become smaller than those observed in the stream when both the number of subhalos and their scale radii are increased. In future work, we plan to investigate joint constraints from stream density and kinematic measurements.

\subsection{Comparison to prior Stream-based Constraints}

\citet{2019ApJ...880...38B} showed that the spur feature can be explained by an interaction with a single compact ($r_s/r_{s\rm{cdm}} \lesssim 0.2$) subhalo with a mass $10^{5.5}-10^8~M_\odot$. The density of the tentative perturber was recently explored in the context of self interacting dark matter (SIDM; \citealt{2000PhRvL..84.3760S}), and is consistent with gravothermal collapse \citep{gd1_core_collapse}. Here we have presented the first analysis of GD-1 in the realistic regime of many subhalo impacts while allowing for the subhalo concentration to vary. Even with more numerous impacts we find that our constraints prefer subhalos that are more concentrated than CDM expectations, and could be explained by SIDM. We can draw a direct comparison between our results and those of \citet{2019ApJ...880...38B, gd1_core_collapse} by considering our Model I, where we vary the dark matter half-mode mass. For $M_{\rm hm} > 10^8~M_\odot$, our models experience only a single direct subhalo impact with a subhalo more massive than $\approx  10^7~M_\odot$. For this $M_{\rm hm}$, we find $r_s/r_{s,\rm{cdm}} < 0.4$ at 68\% confidence, consistent with \citet{2019ApJ...880...38B} and the SIDM scenario \citep{gd1_core_collapse}. 

Our constraints on the number of subhalos in the Milky Way are consistent with \citet{2021JCAP...10..043B}, who model the same GD-1 stream using its on-sky density, though excluding the spur-component. There is a slight preference for more numerous subhalos compared to their work, though our results are in agreement within the 68\% confidence region (Fig.~\ref{fig: shmf}). We also place limits on the dark matter half-mode mass, and find a 1D marginal constraint of $\log_{10}\left(M_{ hm}/M_\odot\right) < 8.6$ at 95\% confidence. This is consistent with $\log_{10}\left(M_{hm}/M_\odot\right) < 8.1$  \citep{2021JCAP...10..043B}. We find a larger upper-limit because we allow subhalos to be more compact in our analysis. Our constraint on $M_{hm}$ provides evidence for the existence of low-mass subhalos below $M\lesssim 10^8~M_\odot$. When extrapolated above $10^9~M_\odot$, our results are  consistent with the number of classical Milky Way satellites \citep{2019MNRAS.487.1380G}. 

While our constraints on the number of subhalos are in agreement with \citet{2021JCAP...10..043B}, we find a preference for more compact subhalos than they report. Specifically, their model fixed $r_s/r_{s,\rm{cdm}} = 1$, whereas we find that subhalos are approximately 60\% more compact than CDM expectations when conditioning on the CDM subhalo number density. This difference may be partly due to the exclusion of the spur in their analysis, since the spur represents the largest surface density fluctuation in the GD-1 stream. In our analysis, removing the spur does not significantly affect the velocity dispersion in the bin $\phi_1 \in [-40, -20]$. If the spur was produced by a subhalo encounter, it is plausible that the central region of the stream would exhibit an elevated velocity dispersion, as we observe, regardless of whether the spur or main stream is selected. However, when considering the stream's density, excluding visually disturbed regions such as the spur could diminish the preference for more compact subhalos that we find. In future work, we will incorporate the surface density of the entire stream into our modeling to better evaluate any potential discrepancy between the stream's kinematics and its on-sky density.

Recently \citet{2025ApJ...989...38C} explored $N-$body simulations of the GD-1 stream in a time-dependent potential. In their work, it was possible to explain the measured dispersion of GD-1 with CDM subhalos. There are a number of differences between our methodologies. First, \citet{2025ApJ...989...38C} has time-dependence in their potential, while our potential model is static. Second, their analysis uses collisionless $N-$body simulations for the dissolving globular cluster, while we use a particle-spray prescription. Third, they consider stream ages up to $10~\rm{Gyr}$, while our maximum dynamical age for the tidal tails is $8~\rm{Gyr}$. It is possible that additional time-dependence in the potential could raise the velocity dispersion of the stream (see, e.g., \citealt{2025arXiv250903599P}), though GD-1's pericenter is only $\approx 14~\rm{kpc}$ \citep{2020ApJ...892L..37B}, so baryonic effects like disk shocking is unlikely to be important for this stream. The stream formation prescription we have adopted \citep{2025ApJS..276...32C} is simplified compared to $N-$body cluster dissolution, though when comparing the velocity dispersions from our model to direct $N-$body simulations we find strong agreement (Appendix~\ref{app: nbody}). We have also tried a 10~\rm{Gyr} model, and still find a preference for more compact subhalos. Our modeling framework is flexible enough to deal with time-dependence in the potential \citep{2025ApJ...983...68N}, so we defer an exploration of this effect to future work. We view this study as constraints on dark matter substructure in an otherwise smooth and static potential model.

\subsection{Comparison to Strong Lensing}
Our constraint on the subhalo mass-concentration relation favors more compact subhalos than the constraints derived from 11 strong gravitational lenses \citep{2020MNRAS.492L..12G} at the 95\% confidence level. At a scale of $M_{200} \approx 3 \times 10^6~M_\odot$, our constraint is consistent with the inferred concentration of the low-mass perturber detected in the JVAS B1938+666 strong lens system \citep{2025NatAs...9.1714P,2026NatAs.tmp....5V}. \citet{2021MNRAS.507.1662M} reported a preference for unusually high subhalo concentrations using the gravitational lens galaxy SDSSJ0946+1006. Depending on the assumed density profile, they found concentrations of approximately $70$ for a subhalo mass of $3\times10^{10}~M_\odot$ and about $1000$ for $5\times10^9~M_\odot$. Our constraints do not require high-mass ($10^9~M_\odot$) subhalos to have anomalously high concentrations in the regime with numerous subhalo impacts (see Fig.~\ref{fig: concentration}). However, if GD-1 was perturbed by a single massive subhalo with a mass of $\sim 10^8~M_\odot$, we find its scale radius must be 20-70\% more compact than CDM expectations. Applying our analysis to additional streams will provide a crucial consistency test, since each Milky Way stream traces the same global density field.

\subsection{Connection to Dark Matter Particle Theories}

We have presented constraints on the mass-concentration relation of dark matter subhalos from $10^5-10^9~M_\odot$. Our constraints can be compared to expectations for distinct dark matter particle theories. In both Models I and II, we find a preference for subhalos more compact than CDM expectations. While we place a limit on the WDM half-mode mass of $M_{hm} < 10^{8.6}~M_\odot$ (95\% confidence), the preference for more compact subhalos than CDM appears in tension with WDM, which typically yields lower characteristic central densities (e.g., \citealt{2001ApJ...556...93B,2014MNRAS.439..300L}) since the free-streaming of WDM particles suppresses small-scale power. However, in WDM prompt cusps \citep{2023MNRAS.522L..78D} can enhance the central density of halos over CDM halos, so our constraints do not necessarily rule out WDM. An exploration of the consistency of our WDM constraint with prompt cusps is beyond the scope of the present work, but marks an interesting future consideration.

Alternative physics such as SIDM can produces more concentrated subhalos, as can Atomic Dark Matter (ADM; \citealt{2010JCAP...05..021K}). Recently, \citet{adm_gemmell} quantified the concentration of ADM subhalos using the statistic $R_{200,m} / R_{1/2}$, where $R_{200,m}$ is the radius of a subhalo enclosing 200 times the mean matter density of the Universe, and $R_{1/2}$ is the radius enclosing half of the subhalo's mass. At $10^7~M_\odot$ their ADM simulations produce subhalos with $R_{200,m} / R_{1/2}$ from $\approx 4$ to $100$, while CDM subhalos have values at or below $\approx 15$. At the same mass scale ($10^7~M_\odot$), our model prefers $R_{200,m} / R_{1/2} \approx 40$, and likely supports higher values as we run into our prior on concentrations. Because our constraints are agnostic to the exact form of dark matter, the output of our analysis can be used to test consistency, or inconsistency, with different dark matter models. This highlights the utility of stellar stream kinematics in constraining dark matter microphysics.

\subsection{Assumptions and Future Directions}\label{sec: assumptions}
We have made a number of simplifying assumptions for the gravitational potential and the internal structure of globular cluster streams. We expand on our assumptions and possible future directions to relax them below.

\begin{itemize}
    \item \textbf{Static potential:} This work presents constraints on dark matter substructure in an otherwise static Milky Way potential. Time-dependence in the potential over several gigayears can lead to additional stream heating that we have not captured in our modeling \citep{2025arXiv250903599P}. We note, however, that GD-1 is on a retrograde orbit with a pericenter of $\approx 13.8~\rm{kpc}$ \citep{2020ApJ...892L..37B}, making the stream less susceptible to baryonic perturbations local to the disk including the galactic bar (e.g., \citealt{2017NatAs...1..633P}). The buildup of the Milky Way halo could still lead to time-dependent orbital effects on the stream, including perturbations due to the Sagittarius Dwarf Spheroidal Galaxy \citep{2022MNRAS.516.1685D}. However, from their estimates of energy dispersion we expect only a modest average increase in the stream's velocity dispersion due to Sagittarius. Still, the perturbative methodology we have used here \citep{2025ApJ...983...68N} is not limited to static and symmetric potentials, and can be extended to time-evolving potentials in future work.

    \item \textbf{Globular cluster dissolution:} Recent work \citep{2025arXiv250915307W} shows that the ejection of stars due to binary interactions can increase the velocity dispersion of tidal tails, particularly for more massive stars. However, the expected contribution of heating from binary interactions alone is insufficient to explain the $5~\mathrm{km/s}$ dispersion we measure in the central region of GD-1. Because our models do not account for binary interactions, it is possible that we have overestimated the contribution of heating from subhalos. Black holes can also increase the central velocity dispersion of globular clusters, but only at the level of $0.2~\mathrm{km/s}$ for a Palomar 5-type cluster \citep{2021NatAs...5..957G}. These results highlight the need for realistic modeling of globular clusters in time-dependent potentials (e.g., \citealt{2025ApJ...989...38C,2025arXiv250903599P}) to improve the use of kinematics as a diagnostic for dark matter substructure.

    \item \textbf{Comparison to CDM at $z=0$:} An advantage of our analysis is its flexibility and empirical nature: subhalos are required to follow a mass function, but the normalization, half-mode mass, and mass--size relation of the subhalos are all allowed to vary. Comparison to CDM and alternative models is therefore a post-processing step. When comparing to CDM, we have used the $z = 0$ distribution of subhalos as a benchmark. Our simulations extend back to $8\,\mathrm{Gyr}$, and the subhalo mass function can evolve significantly over that timescale (e.g., \citealt{2004MNRAS.355..819G,2023MNRAS.523..428B}). Still, the probability of a stream--subhalo interaction is highest closer to the present day, when tidal tails are longest, and substantially lower in the past when the stream is very short. We therefore do not expect significant bias in our assumption. Future work can consider impact times as another dimension to compare against CDM. We also note that the LMC can induce boosts in the number of subhalo interactions, up to a factor of $\approx 2$ \citep{2024ApJ...974..286A,2024arXiv240611989M}. This can impact our comparison to CDM expectations, though our inference of $f_{\rm sub}$, which is independent of CDM assumptions, supports a two-fold increase in the number of subhalo interactions over CDM expectations.

    \item \textbf{The Velocity Dispersion of GD-1:} We have combined radial velocity measurements from several datasets over multiple epochs to constrain the velocity dispersion of the GD-1 stream while also constraining the contribution of the dispersion due to binarity (e.g., \citealt{2026arXiv260306790P}). Future radial velocity measurements will enable a stronger constraint on binarity. We have not considered transverse velocities as a measure of substructure in this work, though the proper motion dispersion in our models is typically higher with subhalos. The improved proper motion precision from {\it Gaia} DR4 will yield an additional constraint on substructure and will provide an independent test of the inference we have presented here. In addition, we have modeled the streams main ridgeline rather than the broader cocoon component. The ridgeline is reproduced in our models, while the cocoon component has higher velocity dispersions consistent with its broader width \citep{2025ApJ...980...71V}, and could be explained by GD-1 forming and accretting with a more massive subhalo of the Milky Way \citep{2019ApJ...881..106M}. A joint analysis of the stream's preprocessing and subsequent evolution in the Milky Way is the subject of future work.

\end{itemize}
 
\section{Summary and Conclusion}\label{sec: summary_and_discuss}
We have presented a means to map the observed velocity dispersion of tidal tails to a constraint on the number and concentration of low-mass dark matter subhalos. In the highest velocity dispersion bin, we find that the velocity dispersion of GD-1 is $\approx 4~\rm{km/s}$ higher than unperturbed models. Thus, the kinematics of GD-1 are inconsistent with its formation in a smooth Milky Way halo. We use perturbation theory to model the stream as a function of subhalo population and internal subhalo parameters \citep{2025ApJ...983...68N}. We find that the radial velocity dispersion of the stream is naturally explained by a population of subhalos accounting for $f_{\mathrm{sub}} = 0.05^{+0.08}_{-0.03}$ of the Milky Way's mass (68\% confidence). This mass fraction is consistent with CDM expectations, though there is a wide range in cosmological simulations ($\sim 5-12\%$, e.g., \citealt{2004MNRAS.355..819G,2008MNRAS.391.1685S,2011MNRAS.410.2309G,2017PhRvD..95f3003S,2023ApJ...945..159N}), and baryons can shift $f_{\rm sub}$ to lower numbers \citep{2012MNRAS.422.1231G,2017MNRAS.471.1709G,2018ApJ...859..129N,2023MNRAS.523..428B}.

Constraints on the mass–size relation of subhalos indicate more compact values than those predicted by CDM. For Model I, we treat $r_s/r_{s,\rm{cdm}}$ as independent of mass, and explore constraints on this parameter as a function of $f_{\rm sub}$. For $f_{\rm sub} = 10\%$, we find $r_{s}/r_{s,\rm{cdm}} = 0.3 \pm^{0.3}_{0.2}$ at 68\% confidence. The 95\% upper limit is $r_{s}/r_{s,\rm{cdm}} < 1$. In Model II we allow the ratio $r_{s}/r_{s,\rm{cdm}}$ to vary as a broken power-law in mass. For this model, we find that there is a preference for more compact subhalos below $10^7~M_\odot$, and CDM concentrations above this mass-scale. Our constraint on the mass-concentration relation for Hernquist subhalos with $M_{200} = [10^5, 10^6, 10^7]~M_\odot$ is $c_{-2} > [422,190,90]$ at 68\% confidence ($c_{-2} > [115,57,27]$ at 95\% confidence). CDM expectations for these masses are $c_{-2} = [50, 43, 37]$, respectively \citep{2017MNRAS.466.4974M}.

Our results suggest a possible deviation from the CDM mass-size relation at low subhalo masses, where subhalos are expected to be completely dark matter dominated without any baryonic component. Our approach can be combined with deep photometry of many streams expected from the Rubin Observatory \citep{ivezic2019}, and more stringent limits from velocity dispersions alone will be within reach with additional data from radial velocity surveys such as DESI \citep{2024AJ....168...58D}, S5 \citep{2019MNRAS.490.3508L}, 4MOST \citep{2019Msngr.175....3D}, WEAVE \citep{2024MNRAS.530.2688J}, and Via (Via collaboration, in prep.). Extending our methodology to additional streams with these surveys will test our results from GD-1, since each stream traces the dark matter density field local to its orbit. Collectively, these datasets will deliver constraints on the abundance and properties of low-mass dark matter subhalos, and ultimately constrain the nature of the dark matter particle.

\section*{Acknowledgements}
JN is supported by a National Science Foundation Graduate Research Fellowship, Grant No. DGE 2039656. Any opinions, findings, and conclusions or recommendations expressed in this material are those of the author(s) and do not necessarily reflect the views of the National Science Foundation. We are pleased to acknowledge that the work reported on in this paper was substantially performed using the Princeton Research Computing resources at Princeton University which is a consortium of groups led by the Princeton Institute for Computational Science and Engineering (PICSciE) and Office of Information Technology’s Research Computing. We are grateful to Sergey Koposov for sharing radial velocity data on LAMOST and SDSS stream members. We thank Mariangela Lisanti, Charlie Conroy, Duncan Adams, Andrew Benson, Robel Geda, Nathaniel Starkman, Kareem El-Badry, Kathryn Johnston, Alex Drlica-Wagner, and Newlin Weatherford for helpful discussions regarding the manuscript. We also thank the anonymous referee for useful comments and suggestions.

\begin{appendix}
\begin{figure}[h!]
\centering\includegraphics[scale=.62]{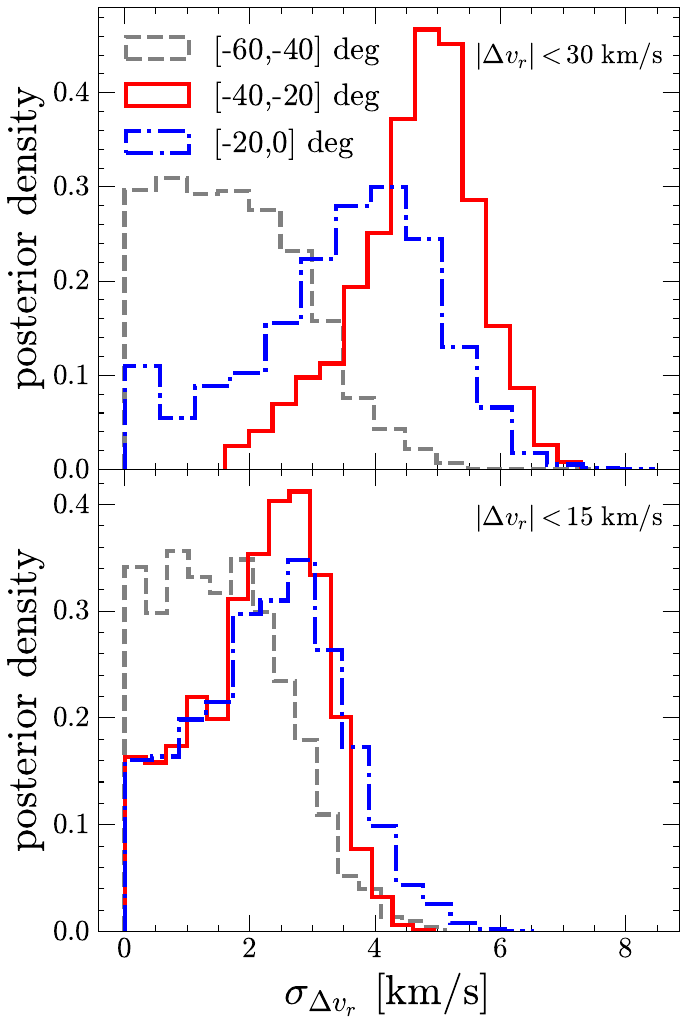}
    \caption{1D marginal posterior densities for the intrinsic velocity dispersion in each of the three bins. }
    \label{fig: sigma_vr_posterior}
\end{figure} 
\section{Intrinsic Velocity Dispersion Posteriors}\label{app: intrinsic_dispersion}
Here we provide constraints on the intrinsic velocity dispersion, $\sigma_{\Delta v_r}$, for two velocity cuts. Posteriors for each bin are provided in Fig.~\ref{fig: sigma_vr_posterior}. The top panel is for $\vert\Delta v_r \vert < 30~\rm{km/s}$, and the bottom uses a tighter selection $\vert\Delta v_r \vert < 15~\rm{km/s}$. At and below the latter cut, we find consistency with \citet{2021ApJ...911L..32G} who reports $\sigma_{\Delta v_r} = 2.3 \pm0.3~\rm{km/s}$ using data from MMT.

\begin{figure}[h]
\centering\includegraphics[scale=.62]{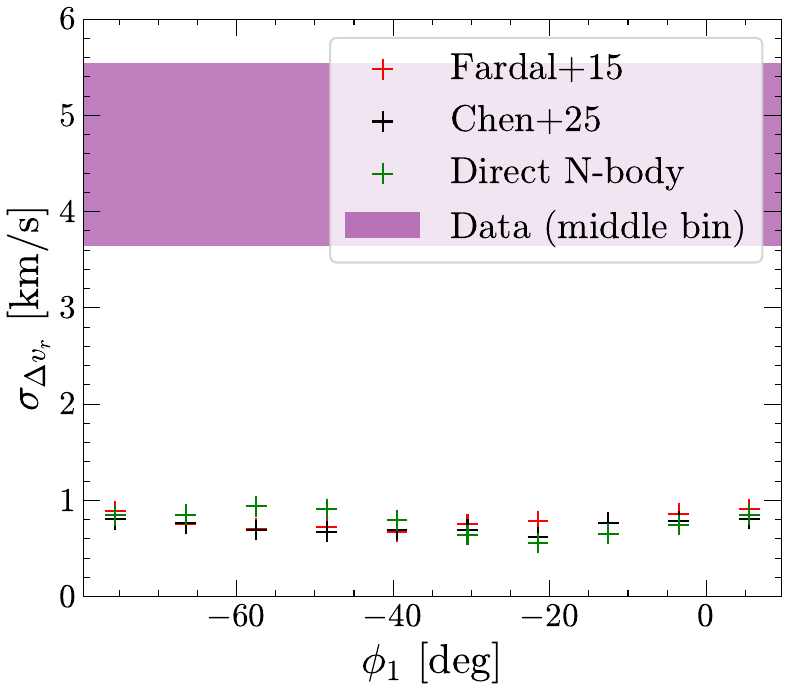}
    \caption{The intrinsic velocity dispersion of three different stream models. Particle spray models from \citet{2015MNRAS.452..301F} and \citet{2025ApJS..276...32C} are shown in red and black, respectively.  The green points are from a direct $N-$body simulation of a GD-1 like stream. The intrinsic dispersion of GD-1 is shown as the purple band. The models show good agreement for the velocity dispersion profile. }
    \label{fig: model_comparison}
\end{figure}

\section{Comparison to $N$-body Simulations}\label{app: nbody}
The radial velocity dispersion profile for the same stream explored with three different models is illustrated in Fig.~\ref{fig: model_comparison}. We consider the particle-spray models from \citet{2015MNRAS.452..301F} and \citet{2025ApJS..276...32C}, and a direct $N-$body model of GD-1 generated using the code \texttt{PeTar} \citep{10.1093/mnras/staa1915}. The intrinsic radial velocity dispersion measured from data is shown as the purple band. The models show excellent agreement.

\section{Posterior for Model II}\label{app: model_II}
Here we provide the posterior distribution for Model II.
\begin{figure*}
\centering\includegraphics[scale=.52]{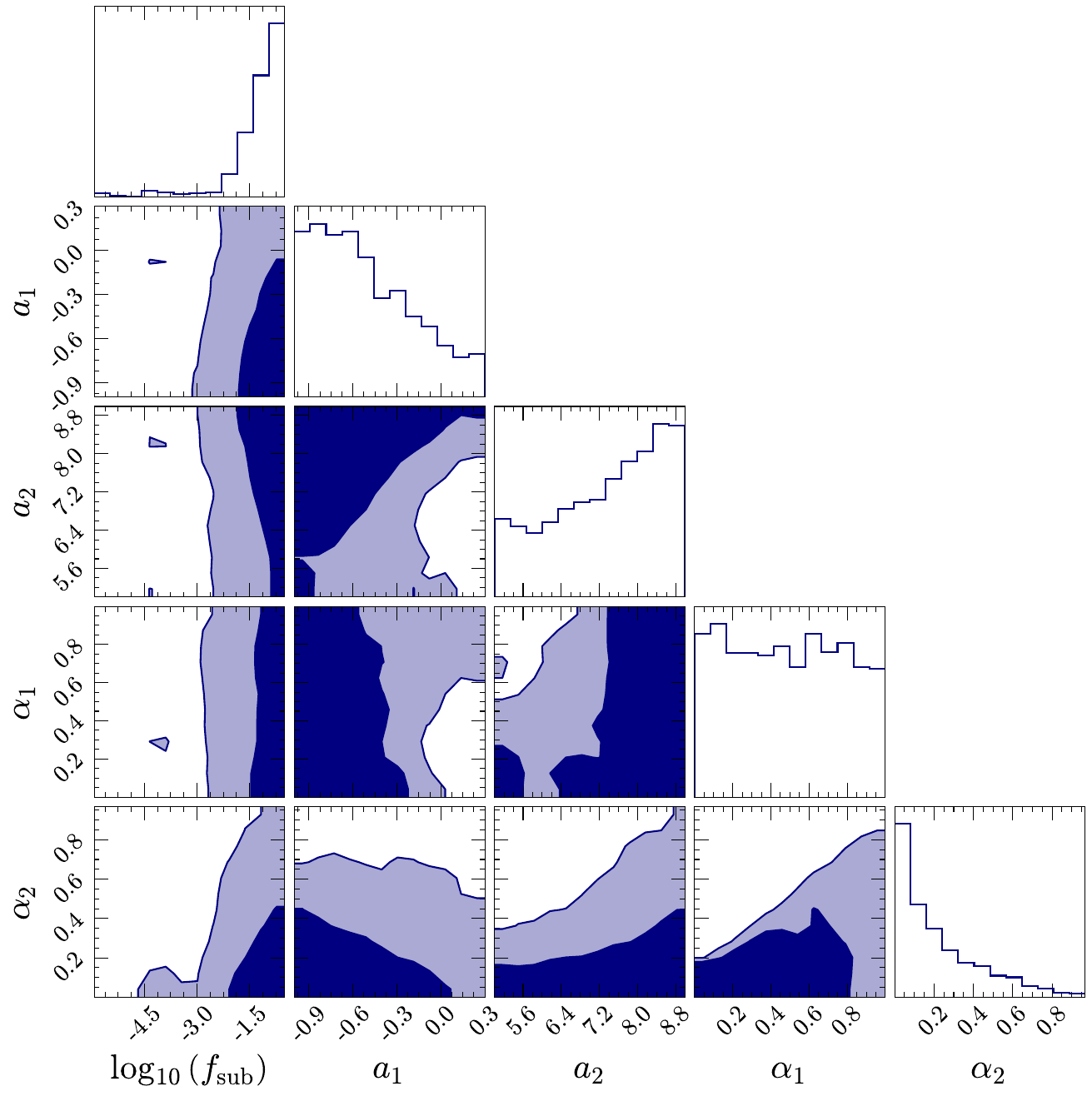}
    \caption{Posterior distribution for Model II. }
    \label{fig: model_II_posterior}
\end{figure*} 
Constraints on the five parameters are plotted in Fig.~\ref{fig: model_II_posterior}.

\clearpage
  
\end{appendix}

\bibliography{thebib}{}
\bibliographystyle{aasjournalv7}
\end{document}